\documentclass[aps,twocolumn,floatfix,showpacs,amssymb]{revtex4}
\usepackage{mathbbol}
\usepackage{color}
\usepackage{hyperref}
\usepackage[utf8]{inputenc}
\usepackage{graphicx,bm}
\usepackage{amsmath}
\usepackage{braket}
\DeclareMathOperator{\arcsinh}{arcsinh}

\newcommand{\Rsep}{R_0}%{R_{\rm sep}}
\newcommand{\mref}{m_{\rm r}}
\newcommand{\Rstars}{{R_s^\star}}
\newcommand{\astars}{{a_s^\star}}

\newcommand{\expos}{s}
\newcommand{\Radius}{R}

\newcommand{\asc}{a}
\newcommand{\stat}{S}

\begin{document}

\title{Universal spectrum of isolated three-body resonances}
\author{Ludovic Pricoupenko}

\affiliation
{
Laboratoire de Physique Th\'{e}orique de la Mati\`{e}re Condens\'{e}e, Sorbonne Universit\'{e},  CNRS UMR 7600, F-75005, Paris, France.
}
\date{\today}

\begin{abstract}
The exact wavefunction of an isolated three-body resonance at finite scattering length is obtained for two identical particles interacting with another one via a pairwise zero-range potential. The corresponding universal spectrum is studied as a function of the scattering length. The universality of the results is illustrated by considering a model with finite-range interactions.
\end{abstract}
\date{\today}
\maketitle

\section{Introduction}

The search for universal solutions to the quantum three-body problem with short range interactions began with the pioneering work of Skorniakov and Ter-Martirosian (STM) on the neutron-deuteron scattering problem \cite{Sko57}. A series of fascinating results followed, culminating in the Efimov effect \cite{Dan61,Min62a,Min62b,Efi70,Efi71}. Universality means that, regardless of the actual interactions of the system that define what can be called a 'reference model', the low-energy properties can be described using a few parameters. The reference model can therefore be replaced by a contact model defined by zero-range interactions, where for a given energy, the wavefunction is set by these few parameters. Universality happens when the dominant interaction between particles is close to the unitarity of two-body s-wave scattering, such as the neutron-neutron interaction and has led to recent predictions in nuclear physics \cite{Ham21}. Universality is also an important line of research in ultracold physics, which offers the possibility of tuning the scattering length and reaching unitarity with Feshbach resonances \cite{Ino98,Chi10}. This led to the first observation of the Efimov effect \cite{Kra06}. Alongside dramatic experimental progress and a detailed analysis of the Efimov effect \cite{Nai17}, one can point to remarkable theoretical results in the three-dimensional case~\cite{Bed00,Pet03,Pet04,Wer06a,Kar07a,Kar07b,Gog08}. 
 
Apart from the Efimov effect, which has been extensively studied, Isolated Three-Body Resonances (ITBR) close to unitarity have received less attention despite their universal character. The term 'isolated' resonance refers to the fact that when the resonance occurs at the three-body threshold for a given scattering length, the spectrum reduces to an isolated point, whilst the Efimov effect at unitarity is characterized by an accumulation point at zero energy. The universality follows from the scale invariance of the wavefunction in a domain of short distances. The associated scaling exponent denoted in what follows by ${\expos}$, depends only on the configuration (angular momentum, quantum statistics and mass ratio). The Efimov effect occurs when ${\expos}$ is imaginary   whatever the short range details of the actual interactions with an infinite number of states at unitarity. In deep contrast, an ITBR may occur in all other configurations where the scaling exponent is real and positive but only for specific interactions \cite{Nis08,Saf13} moreover there is only one shallow state at unitarity. From this perspective, Efimov states and ITBR encompass all possible universal physics for Borromean systems with short-range s-wave resonant interactions.  A first result was found at unitarity and for small values of the scaling exponent ${\expos}$ with a one-parameter spectrum law in Refs.~\cite{Wer06b,Gao15}. For arbitrary ${\expos}$, it has been shown that the spectrum at unitarity is defined by two three-body parameters, {whereas  the Efimov spectrum is a one parameter law}~\cite{Pri23,Pri24}.

In the present work, the universal spectrum of ITBR is derived as a function of the scattering length ${\asc}$ for three types of systems in configurations without Efimov effect. They can consist of two identical particles (mass ${M}$) each interacting  with an impurity of mass ${m}$ close to a s-wave resonance (i.e. the particle-impurity scattering length ${\asc}$ is large). They are denoted as 2FI systems when the two identical particles are fermions and 2BI systems when they are bosons (with a negligible mutual $s$ wave scattering length). They can also consist of three identical interacting bosons (denoted as 3B systems) of mass ${M}$ close to a s-wave resonance but with a non-zero total orbital angular momentum. The main results are as follows: ${i)}$  ITBR states are expressed in terms of universal functions obtained from the exact solutions of a Lippmann Schwinger-like equation ; ${ii)}$ in the limit of small detuning, the spectrum is obtained analytically for ${\expos \gtrsim 1.5}$; otherwise, it is deduced from an equation where only one universal parameter is evaluated numerically; ${iii)}$ for ${\asc>0}$, the modification of the  Kartavtsev Malyckh (KM) regular states of Refs.~\cite{Kar07a,Kar07b}, due to an ITBR is discussed; ${iv)}$ the universality is tested successfully in an  example of ITBR, illustrating also a way to evaluate the three-body parameters from a given spectrum.

\section{Contact model}

The particles of mass ${M}$ labelled by $1$ or $2$ and the particle of mass $m$ labelled by $3$, are considered in the center of mass frame. The position and  wavevector of the particle ${i}$ are ${\mathbf r_i}$ and ${\mathbf k_i}$. There are three sets of Jacobi coordinates ${\boldsymbol \rho=(\boldsymbol \rho^{(ij)}_1,\boldsymbol \rho^{(ij)}_2)}$, each associated with a pair $(ij)$. Using the reference mass ${\mref=mM/(M+m)}$, the coordinates associated with the pair $(23)$ are {the relative coordinates} ${\boldsymbol \rho^{(23)}_1=\mathbf r_{23}}$ {and  the position of the particle 1 with respect to this pair}  ${\boldsymbol \rho^{(23)}_2=  \mathbf r_{23} \tan \theta- \mathbf r_{13} \sec \theta}$  where ${\mathbf r_{ij}=\mathbf r_j-\mathbf r_i}$ and ${\theta =\arcsin(\mref/m)}$. The coordinates ${(\boldsymbol \rho^{(12)}_1,\boldsymbol \rho^{(12)}_2)}$ [or ${(\boldsymbol \rho^{(31)}_1,\boldsymbol \rho^{(31)}_2)}$] are deduced from ${(\boldsymbol \rho^{(23)}_1,\boldsymbol \rho^{(23)}_2)}$ by using a ${2\times2}$ active rotation matrix of angle ${\frac{3\pi}{4}-\frac{\theta}{2}}$ [or ${(-\frac{\pi}{2}-\theta)}$]. The hyperradius ${\rho=\|\boldsymbol \rho\|}$ {that measures the relative distance between the three particles}, is thus the same for the three sets. One has also the properties ${\boldsymbol \rho^{(ij)}_1=-\boldsymbol \rho^{(ji)}_1}$ and ${\boldsymbol \rho^{(ji)}_2=\boldsymbol \rho^{(ij)}_2}$. More details on the Jacobi coordinates are given in Appendix \ref{app:Jacobi}. The interaction between two particles ${(ij)}$ is modeled by the Wigner-Bethe-Peierls contact condition when ${\rho^{(ij)}_1\to 0}$ for a given ${\boldsymbol \rho^{(ij)}_2}$:  
\begin{equation}
\langle \boldsymbol \rho | \Psi \rangle =  \langle \boldsymbol \rho^{(ij)}_2 | A_{ij} \rangle
\left(\frac{1}{a} -\frac{1}{\rho^{(ij)}_1} \right) 
 + O(\rho^{(ij)}_1) .
\label{eq:contact_ij}
\end{equation} 
It is convenient to introduce a statistical factor ${\stat}$ related to the exchange symmetry: ${|A_{12}\rangle=0}$, ${|A_{23}\rangle=-|A_{13}\rangle}$ with ${\stat=-1}$ for the 2FI system; ${|A_{12}\rangle=0}$, ${|A_{13}\rangle= |A_{23}\rangle}$  with ${\stat=1}$ for the 2BI system and ${|A_{12}\rangle=|A_{13}\rangle= |A_{23}\rangle}$ with ${\stat=2}$ for the 3B system. Isotropy allows one to isolate each component of angular momentum ${\ell \hbar}$:
\begin{equation}
\langle \boldsymbol \rho_2^{(23)} | A_{23} \rangle= \langle  \rho_2^{(23)} | \mathcal A_\ell \rangle 
P_\ell(\hat{\mathbf e}_{z}\cdot\boldsymbol \rho_2^{(23)}/ \rho_2^{(23)}) .
\end{equation}
At unitarity (${|\asc|=\infty}$),  the contact conditions  \eqref{eq:contact_ij} is scale invariant so that the wavefunction is separable \cite{Wer06a}:
\begin{equation}
\langle \boldsymbol \rho | \Psi \rangle {=} \frac{\langle \rho | \mathcal A_\ell\rangle}{\rho} \Phi_{\ell}\left(\frac{\boldsymbol \rho}{\rho}\right) ,
\label{eq:Psi_separable}
\end{equation}
where ${\Phi_{\ell}(\boldsymbol \rho/\rho)}$ is {a normalized} eigenstate of the Laplacian on the unit hypersphere with the orbital momentum ${\ell \hbar}$. The contact conditions \eqref{eq:contact_ij}  gives the eigenvalues ${4-\gamma^2}$. {The possible values of ${\gamma}$ are functions of the mass ratio ${M/m}$ (or of ${\theta}$) and of ${\ell}$ \cite{Wer06a}.} Using this solution in the Schr\"{o}dinger equation for ${\langle \boldsymbol \rho|\Psi \rangle}$, one finds that ${\rho \langle \rho | \mathcal A_\ell\rangle}$ satisfies a radial 2D Schr\"{o}dinger equation with an effective centrifugal barrier ${\hbar^2 \gamma^2/(2\mref\rho^2)}$. For given ${\stat}$ and ${M/m}$, an ITBR can exist only for ${\ell}$ such that all the eigenvalues ${\gamma^2}$ are positive. The smallest value of  ${|\gamma|}$ is the scaling exponent $\expos$ which corresponds to the lowest centrifugal  barrier.  In a contact model, an ITBR is then characterized by the singularity ${\langle \rho | \mathcal A_\ell\rangle \propto \rho^{-1-\expos}}$ when ${\rho\to 0}$. At unitarity the hyperradial function for a bound state of energy ${E=-{\hbar^2q^2}/{(2\mref)}}$ is
\begin{equation}
 \langle \rho | \mathcal A_\ell \rangle \propto \frac{K_\expos( q \rho )}{\rho}  .
 \label{eq:Fradial}
\end{equation}
At finite scattering length, the function  ${\rho \langle \rho | \mathcal A_\ell\rangle}$ still satisfies this 2D Schr\"{o}dinger equation  but only for ${\rho\ll |a|}$. In physical systems, an ITBR follows from the existence of actual attractive interactions that truncate the centrifugal barrier in a region of short hyperradius  ${\rho<\Rsep}$. The length ${\Rsep}$ is defined such that for interparticle distances larger than ${\Rsep}$, the wavefunction of the reference model, ${\langle \boldsymbol \rho|\Psi^{\rm ref}\rangle}$ almost coincides with ${\langle \boldsymbol \rho|\Psi \rangle}$. Hence, in the separability region ${\Rsep {\lesssim} \rho \ll |\asc|}$, ${\langle \boldsymbol \rho|\Psi^{\rm ref}\rangle}$ factorizes as in Eq.~\eqref{eq:Psi_separable} with ${\langle \rho |\mathcal A_\ell^{\rm ref} \rangle \simeq \langle \rho |\mathcal A_\ell \rangle}$. 

To take advantage of translational invariance, {it is convenient to consider the reciprocal space} with the wavevectors ${(\boldsymbol \kappa^{(ij)}_1,\boldsymbol \kappa^{(ij)}_2)}$ that conjugate ${(\boldsymbol \rho^{(ij)}_1,\boldsymbol \rho^{(ij)}_2)}$:
\begin{equation}
\boldsymbol \kappa^{(ij)}_1 \cdot \boldsymbol \rho^{(ij)}_1 + \boldsymbol \kappa^{(ij)}_2 \cdot \boldsymbol \rho^{(ij)}_2 = \mathbf k_1\cdot \mathbf r_1+\mathbf k_2\cdot \mathbf r_2+\mathbf k_3\cdot \mathbf r_3 .
\label{eq:k.r}
\end{equation}
{More details about these wavevectors are given in Appendix \ref{app:Jacobi}. One obtains the function ${\langle k | \mathcal A_\ell \rangle}$ from ${\langle \rho | \mathcal A_\ell \rangle}$ in terms of the Hankel transform:
\begin{equation}
\langle k | \mathcal A_\ell \rangle = \int_0^\infty \rho^2 j_\ell(k\rho) \langle \rho |\mathcal A_\ell \rangle d\rho .  
\label{eq:Hankel_transform} 
\end{equation}
Notice that in the reciprocal space, the reference state ${| A^{\rm ref}_{23}   \rangle}$ analogous to ${|A_{23}  \rangle}$ which is associated with the stationary state ${|\Psi^{\rm ref}\rangle}$ of the reference model is defined by
\begin{equation}
\langle \boldsymbol \kappa_2^{(23)} | A^{\rm ref}_{23}  \rangle= \frac{\mref  }{2\pi \hbar^2} \lim_{\kappa_{1}^{(23)}\to 0}
 \langle \boldsymbol \kappa_1^{(23)}, \boldsymbol \kappa_2^{(23)}|V_{23}| \Psi^{\rm ref} \rangle 
\label{eq:Aref}
\end{equation}
where $V_{23}$ is the actual interaction in the reference model between particles $2$ and $3$. For an angular momentum $\ell$, one has
\begin{equation}
\langle \mathbf k | A_{23}^{\rm ref} \rangle= \langle  k | \mathcal A_\ell^{\rm ref} \rangle 
P_\ell(\hat{\mathbf e}_{z} \cdot \mathbf k/k) .
\end{equation}
}

\section{Universal functions solutions of the STM equation}

For a bound state of binding wavenumber $q$, the {two-body} contact conditions in Eq.~\eqref{eq:contact_ij} lead to the STM equation:
\begin{equation}
\langle k | \mathcal L_{\ell,q}^\infty |\mathcal A_\ell\rangle =  \frac{\langle k | \mathcal A_\ell \rangle}{\asc}, 
\label{eq:STM}
\end{equation}
where ${\mathcal L_{\ell,q}^\infty}$ is the following operator
\begin{multline}
 \langle k | \mathcal L_{\ell,q}^\Lambda|\mathcal A_\ell\rangle =   \sqrt{k^2+q^2}  \langle k | \mathcal A_\ell\rangle - \frac{2\stat(-1)^\ell}{\pi \sin (2 \theta)}
\\
\times \int_0^\Lambda \frac{k' dk'}{k} 
Q_\ell\left(\frac{k^2 + k'\,^2+q^2 \cos^2 \theta}{2 k k' \sin \theta}\right)  \langle k' | \mathcal A_\ell\rangle ,
\label{eq:STM_operator}
\end{multline}
considered at ${\Lambda=\infty}$. A derivation of the STM equation is given in Appendix \ref{app:STM}. It is convenient to introduce the reduced wavenumber ${z={k}/{q}}$ and the dimensionless operator ${\mathcal L^\Lambda}$ obtained by substituting ${q \to 1}$ and ${(k,k') \to (z,z')}$ in Eq.~\eqref{eq:STM_operator}. The solutions of Eq.~\eqref{eq:STM} can be then expressed in terms of universal functions ${\langle z | \phi_{\ell,\expos,\tau}\rangle\equiv \langle k | \mathcal A_\ell \rangle}$ that satisfy
\begin{equation}
\langle z | \mathcal L^\infty -\tau |\phi_{\ell,\expos,\tau} \rangle =  0  \quad \text{with} \quad \tau=\frac{1}{qa}  <1 .
 \label{eq:STM_vp}
\end{equation}
The asymptotic behavior  ${\langle z|\phi_{\ell,\expos,\tau} \rangle=O (z^{-2+\expos})}$ when ${z\to \infty}$ is obtained by considering the zero energy limit at unitarity where Eq.~\eqref{eq:STM} is scale invariant or also from the behavior of ${\rho \langle \rho | \mathcal A_l \rangle}$ when ${\rho \to 0}$.  At unitarity ${(\tau=0)}$, the universal functions are obtained directly from Eqs.~(\ref{eq:Psi_separable},\ref{eq:Fradial}) and
{by using the Hankel transform in Eq.~\eqref{eq:Hankel_transform}}. They are given in Appendix \ref{app:phi_unitary}. Importantly, no ITBR {can be} found when a cut-off is introduced in the integral of Eq.~\eqref{eq:STM_vp} \cite{End12}. This is explained by the underlying centrifugal barrier, which prevents Borromean states from being found if a singularity is not somehow imposed in Eq.~\eqref{eq:STM_vp} when ${z\to\infty}$. Instead, standard calculations are only able to recover the KM states  which are regular at the three-body contact. They exist for a sufficiently small value of ${\expos}$ and for ${a>0}$, i.e. in the presence of the shallow dimer of energy ${-\hbar^2/(2\mref \asc^2)}$~\cite{Kar07a,Kar07b}. 

The universal functions can be deduced from their expressions at unitarity via a Lippmann Schwinger-like equation~\cite{limit}:
\begin{equation}
 |\phi_{\ell,\expos,\tau} \rangle = |\phi_{\ell,\expos,0} \rangle + \lim_{\Lambda \to \infty}
\frac{\tau}{\mathcal L^{\Lambda} -\tau}  |\phi_{\ell,\expos,0} \rangle .
\label{eq:vp_inverted}
\end{equation}
This latter equation follows from the asymptotic equivalence ${\langle z|\phi_{\ell,\expos,\tau}\rangle \simeq \langle z|\phi_{\ell,\expos,0}\rangle}$ for ${z\gg1}$ which implies
\begin{equation}
\langle z| \mathcal L^\infty - \mathcal L^\Lambda |\phi_{\ell,\expos,\tau}\rangle \underset{\Lambda\gg 1,z}{\simeq} \langle z | \mathcal L^\infty - \mathcal L^\Lambda  | \phi_{\ell,\expos,0}\rangle ,
\label{eq:approx}
\end{equation} 
so that Eq.~\eqref{eq:STM_vp} can be written as:
\begin{equation}
(\mathcal L^\Lambda-\tau) |\phi_{\ell,\expos,\tau} \rangle \underset{\Lambda\gg1}{=}  \mathcal L^\Lambda  |\phi_{\ell,\expos,0} \rangle 
\label{eq:vp_reg}
\end{equation}
which gives Eq.~\eqref{eq:vp_inverted}.

\section{Universal spectrum}

The spectrum can be found by using a quite general energy dependent log-derivative condition: 
\begin{equation}
\frac{\partial_\rho[\rho \langle \rho | \mathcal A_\ell \rangle]}{\langle \rho | \mathcal A_\ell \rangle} \biggr|_{\rho=\Radius} =-\upsilon+\upsilon' q^2 \Radius^2,
\label{eq:log-derivative}
\end{equation}
where  ${\Radius}$ { satisfies the condition ${q\Radius \ll 1}$. The calculations are performed at a linear order in energy, which is mandatory when ${\expos>1}$, as shown in Ref.~\cite{Pri23}. It is therefore natural to introduce the coefficient ${\upsilon'}$ in the right hand side of Eq.~\eqref{eq:log-derivative} as done in Ref.~\cite{Pri24}. It will be shown in sec. \ref{sec:best-choice} that  in standard situations, the 'best' choice of the parameters is obtained when ${\upsilon'=0}$, so that the two relevant three-body parameters are ${(\Radius,\upsilon)}$}. The quantity ${\expos-\upsilon}$ is the detuning from the threshold : {this means that at unitarity, ${q=0}$ when this detuning is zero.} Now consider the expansion:
\begin{equation}
\rho \langle \rho | \mathcal A_\ell \rangle= \sum_{n=0}^\infty c_{\ell,-\expos,\tau}^{(n)} (q\rho)^{n-\expos} - {r}_{\ell,\expos,\tau} \sum_{n=0}^\infty c_{\ell,\expos,\tau}^{(n)} (q\rho)^{n+\expos} 
\label{eq:expansion_A}
\end{equation}
with the choice of normalization ${c_{\ell,\pm\expos,\tau}^{(0)}=1}$. {This series expansion is just a generalization of the expansion of the modified Bessel function ${K_\expos(q\rho)}$ considered in the vicinity of ${\rho=0}$.} The two sums in the right-hand-side of Eq.~\eqref{eq:expansion_A} are the two independent solutions of Eq.~\eqref{eq:STM_vp} when considered in the configuration space. The coefficient ${r_{\ell,\expos,\tau}}$ is such that ${\lim_{\rho \to \infty} \langle \rho | \mathcal A_\ell\rangle=0}$. {It fixes the balance between the irregular and regular series in Eq.~\eqref{eq:expansion_A} and will be denoted as the balance coefficient}.The coefficients ${c_{\ell,\gamma,\tau}^{(n)}}$ in Eq.~\eqref{eq:expansion_A} are obtained by recurrence by injecting the Hankel transform \eqref{eq:Hankel_transform} of Eq.~\eqref{eq:expansion_A} in Eq.~\eqref{eq:STM}. {At this stage, it is important  to have in mind that 
in  the contact model, ${\langle \rho | \mathcal A_\ell\rangle}$ coincides with the hyperradial solution of a reference model only  asymptotically for ${\rho \gg \Radius}$.} In the low energy limit, Eq.~(\ref{eq:expansion_A}) is truncated at the order ${n=2}$: an approximation compatible with the  dependence of the right-hand-side of Eq.~\eqref{eq:log-derivative} on the energy. At this order, Eq.~\eqref{eq:log-derivative}  gives:
\begin{equation}
\frac{1}{\astars} -q^2 \Rstars = \frac{\expos \Gamma(\expos)^2 r_{\ell,\expos,\tau} }{\pi \Radius}
\left(4 q^2\Radius^2\right)^\expos,
\label{eq:energy-condition}
\end{equation}
where the generalized scattering length ${\astars}$ and the range parameter ${\Rstars}$ are defined by: 
\begin{align}
&\astars=\frac{\pi \Radius}{\expos 4^\expos \Gamma(\expos)^2} 
\left[ 
\frac{\upsilon+\expos+\sum_{i=1}^2\frac{\Radius^i}{\asc^i} \left(\upsilon+\expos+i\right) \frac{dc_{\ell,\expos,\tau}^{(i)}}{d(\tau^i)}}
{\expos \leftrightarrow -\expos}
\right]\label{eq:astars}\\
&\Rstars=\frac{\expos\Radius 4^\expos(\expos-\upsilon) \Gamma(\expos)^2}{\pi(\expos+\upsilon)} \left[
\frac{\upsilon'+\frac{2+\upsilon-\expos}{4(\expos-1)}}{\expos-\upsilon} - {\expos \leftrightarrow -\expos}
\right] .
\label{eq:Rstars}
\end{align}
The coefficients ${c_{\ell,\expos,\tau}^{(i)}}$ are given for ${i=1,2}$ in Eqs.~(\ref{eq:c1},\ref{eq:c2}) {of the appendix \ref{app:coef_expansion}}. Terms of order ${(q\Radius)^3}$ are not included, meaning that for a binding energy ${E<0}$, except if {the balance coefficient} ${r_{\ell,\expos,\tau}}$ is anomalously large, the right-hand-side of Eq.~\eqref{eq:energy-condition} is relevant only for values of the scaling exponent ${\expos}$ smaller than ${\frac{3}{2}}$ \cite{Large_ratio}.  {When one considers the case where  ${\upsilon'=0}$, in the limit of small detuning ${\upsilon \to \expos}$ and also near the unitarity ${\Radius /|\asc|\ll 1}$, the expression of  $\astars$ can be simplified with:
\begin{equation}
\frac{1}{\astars} \simeq \frac{\expos 4^\expos \Gamma(\expos)^2} {\pi (\upsilon+\expos)}
\left(\frac{\upsilon-\expos}{\Radius } + \frac{c_{\ell,-\expos,\tau}^{(1)}}{\asc \tau}\right)
\label{eq:approx_as}
\end{equation}
and for $\Rstars$ when $\expos>1$:
\begin{equation}
\Rstars\simeq \frac{\Radius 4^{\expos-1} \Gamma(\expos)^2}{\pi(\expos-1)} 
\label{eq:approx_Rs}.
\end{equation}
Equations.~(\ref{eq:astars},\ref{eq:Rstars}) have been obtained from calculations performed up to the first order in energy. This approach is therefore similar to what is done in the effective range approximation for the two-body problem. However, this analogy is only formal and the approximation which leads to Eq.~\eqref{eq:energy-condition}, valid in the limit of small three-body energy, has nothing to do with the actual effective range approximation performed at the two-body problem. Indeed, taking into account the effective range at the two-body level for two resonant particles modifies only the STM equation, leading to an additional term ${r_{\rm e} (k^2+q^2)/2}$ in the STM operator of Eq.~\eqref{eq:STM_operator} or equivalently, an additive term ${q r_{\rm e} (z^2+1)/2}$ in the dimensionless operator ${\mathcal L^\Lambda}$ (see for instance Ref.~\cite{Pri13} with the substitution ${R^\star=-r_{\rm e}/2}$). It is assumed in the present work that the effective range is of the order of the potential radius so that this term can be neglected for ${k  r_{\rm e}\ll 1}$. The solutions of this modified STM equation differ from the universal functions when the wavenumber ${k}$ is larger than or of the order of  ${1/ r_{\rm e}}$, a region where also the universal functions differ from the reference functions ${\langle k |\mathcal A^{\rm ref}_\ell \rangle}$. The log-derivative condition is in general considered at an hyperradius $\Radius$ where also these functions differs. Nevertheless, this is not a problem because the log-derivative condition is imposed to give the correct value of the binding wavenumber ${q}$, but  the reference function coincides with the reference function only for small wavenumbers ${k}$ ${(k\Radius \ll1)}$ and absolutely not when ${k}$ is of the order of ${1/r_{\rm e}}$. For such large values of the wavenumber the interaction potential is not negligible in the Schr\"{o}dinger equation and the STM equation cannot model the actual physics.} 

Solutions of Eq.~\eqref{eq:energy-condition} are detailed in the following lines. The pure KM states labelled by ${n=1,2\dots}$ are regular solutions of Eq.~\eqref{eq:STM_vp} i.e. ${r_{\ell,\expos,\tau}=\infty}$, corresponding to specific values of ${\tau=\tau_{n}}$. Their number depends on the configuration studied. Even if the following reasoning is generic, in practice it can be implemented only for a sufficiently small value of the scaling exponent such that ${r_{\ell,\expos,\tau}}$ can be evaluated precisely. Each KM state ${(n)}$ is associated with an universal branch (it depends solely on the mass ratio and the statistics) of the coefficient ${r_{\ell,\expos,\tau}}$ considered as a function of ${\tau}$, with a vertical and positive asymptote located at ${\tau=\tau_{n}}$. For a given branch and an arbitrarily large and positive scattering length, the pure KM state is found from Eq.~\eqref{eq:energy-condition} with ${q=1/(\asc \tau_{n})}$. For a decreasing value of ${\asc>0}$, the KM spectrum deviates from this last law with a solution of Eq.~\eqref{eq:energy-condition} corresponding to a larger values of ${\tau}$ on the same branch of ${r_{\ell,\expos,\tau}}$. Importantly, the spectrum of the shallowest KM state has an endpoint at the particle-dimer continuum threshold if Eq.~\eqref{eq:energy-condition} has a solution at ${q=1/\asc}$.

Let's move on to the analysis of the ITBR. At unitarity, the resonance threshold ${(E=0)}$ is at ${\upsilon=\expos}$ and a shallow bound state exists when ${\expos-\upsilon}$ is small and negative \cite{Pri23,Pri24}. The spectrum is obtained from Eq.~\eqref{eq:energy-condition} by considering the branch of ${r_{\ell,\expos,\tau}}$ starting from ${\tau=-\infty}$ corresponding to the endpoint of the spectrum at the three particles continuum for the scattering length ${\asc=a^{-}}$. {In this limit,  ${r_{\ell,\expos,\tau}=B (-\tau)^{2\expos}}$, where $B$ is a coefficient that can be obtained numerically (see the end of the Appendix \ref{app:coef_expansion}). The position of the endpoint satisfies the approximate equation:
\begin{equation}
\upsilon-\expos +\frac{\Radius}{a^{-}} \frac{c_{\ell,-s,\tau}^{(1)}}{\tau}  =(\upsilon+\expos) B \left(\frac{-\Radius}{a^{-}}\right)^{2\expos}
\end{equation}
which becomes exact in the limit of a small detuning. In this limit, one obtains
\begin{equation}
a^{-} \sim
\left\{
\begin{array}{ll}
  \frac{\Radius}{\expos-\upsilon}  \frac{c_{\ell,-s,\tau}^{(1)}}{\tau} & \text{for} \quad \expos>1/2\\
 -\Radius  \left( \frac{\upsilon+\expos}{\upsilon-\expos } B \right)^{\frac{1}{2\expos}} & \text{for} \quad \expos<1/2
\end{array}
\right.
\label{eq:endpoint_approx}
\end{equation}
}
Let us consider the regime ${\expos<3/2}$. 
\begin{figure}[h]
\centering
\includegraphics[width=8cm]{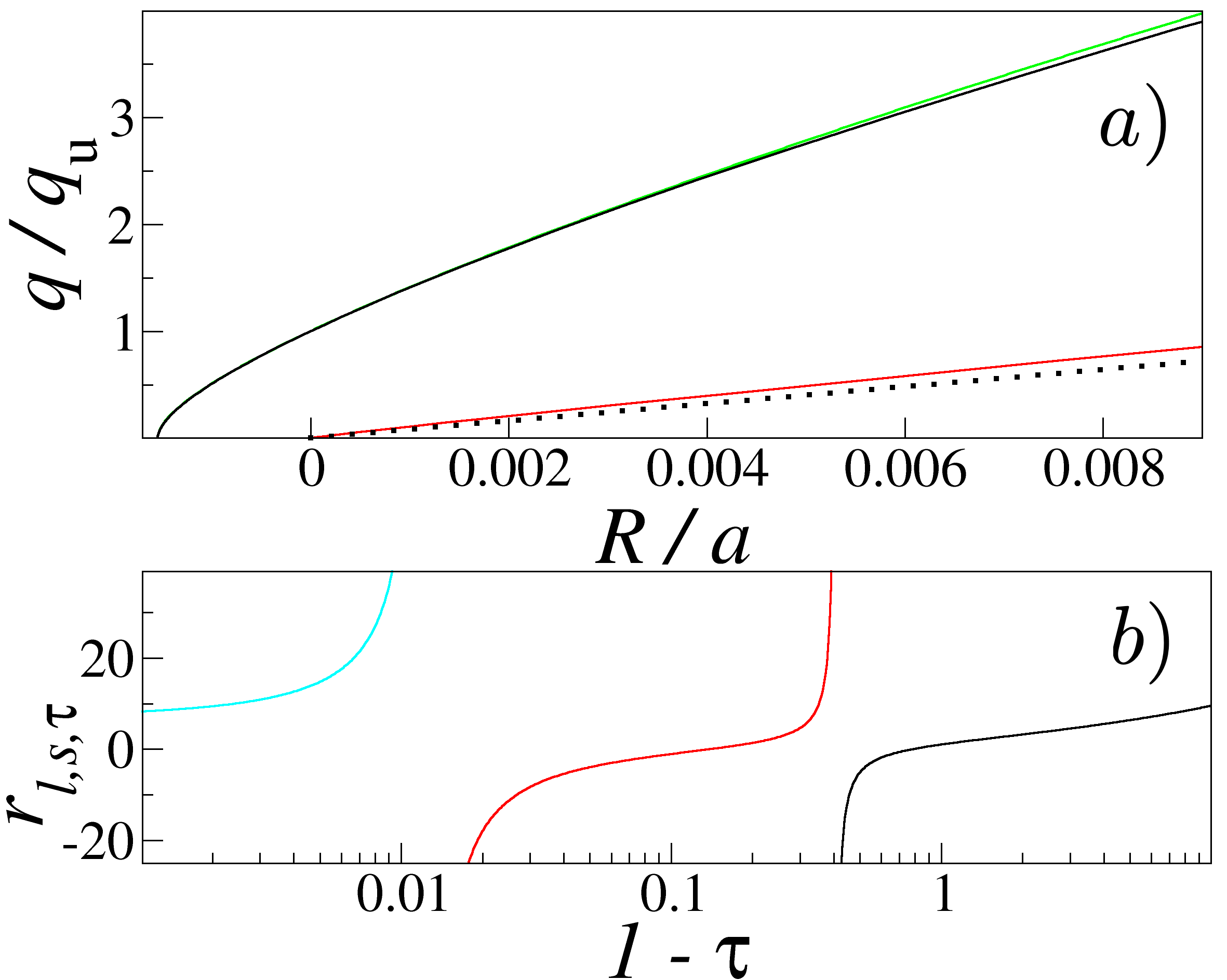}
\caption{(a) Spectrum of a 2FI system with ${\ell=1}$, ${\expos=0.25}$, ${\expos-\upsilon=-5.960\dots\times10^{-2}}$ and ${\upsilon'=0}$. Black solid line: ITBR; red solid line: deepest KM state; dotted line: dimer spectrum; green solid line: example of ITBR of the separable model of {Eqs.~(\ref{eq:V_separable},\ref{eq:form-factor}) for ${\alpha=0}$ (see sec.\ref{sec:example_ITBR_0})}. (b) {Plot of the balance coefficient ${r_{\ell,\expos,\tau}}$ defined in the series of Eq.~\eqref{eq:expansion_A}}.}
\label{fig:1}
\end{figure}
The 2FI system with ${\ell=1}$ is an interesting configuration to illustrate this last discussion. In this system, the Efimov effect occurs for ${M/m >x^{\rm E}=13.60696\dots}$ and there are at most two KM states when ${\expos<1.04 \dots}$: the ${n=1}$ KM state exists for ${x^{\rm E} >M/m>8.17\dots}$ and the ${n=2}$ KM state, for ${x^{\rm E}>M/m>12.9\dots}$ Figure~(\ref{fig:1}a) gives an example of spectrum {plotted} in the unit of the binding wavenumber at unitarity ${q_{\rm u}}$, where  the values of ${(\expos,\Radius,\upsilon,\upsilon')}$ are chosen for comparison with an ITBR in a reference model that will be introduced in sec.~\ref{sec:example_ITBR_0}. The coefficient ${r_{\ell,\expos,\tau}}$ plotted in Fig.~\ref{fig:1}b has three branches and there are thus two KM states. The red branch in the spectrum corresponds to the deepest KM state (associated with the red branch of ${r_{\ell,\expos,\tau}}$). The excited KM state associated with the cyan branch, is almost degenerate with the dimer spectrum. 

\begin{figure}
\centering
\includegraphics[width=8cm]{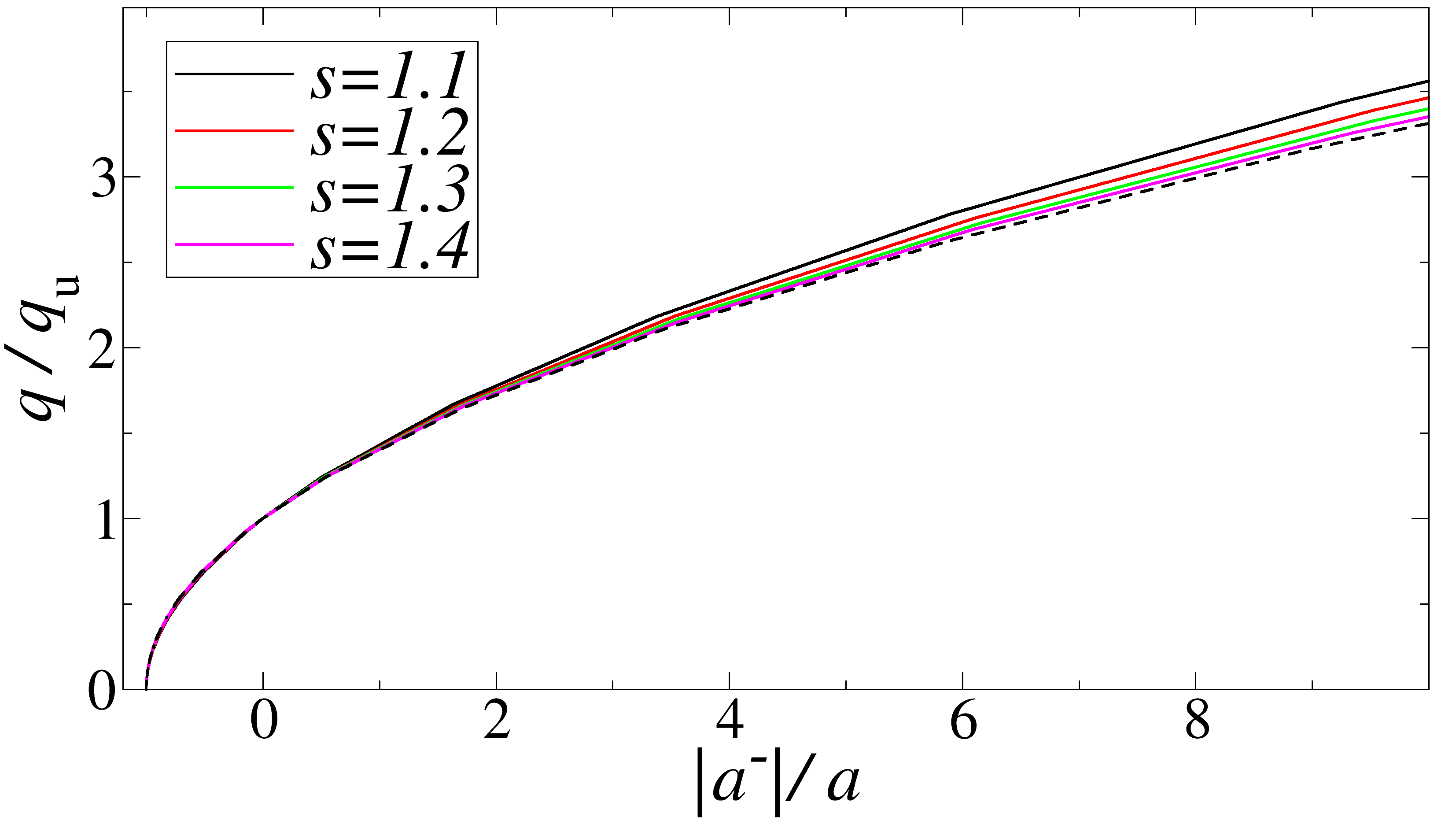}
\caption{Plot of the universal spectrum for ${\upsilon=\expos+10^{-3}}$, ${\upsilon'=0}$ and for increasing values of ${\expos}$ (Solid line). Black dashed line: asymptotic law in Eq.~\eqref{eq:E_s>1} for the same values of ${\expos}$.}
\label{fig:spectrum_expos}
 \end{figure}

Let's return to the general discussion when ${\expos>3/2}$ and the right-hand-side of Eq.~\eqref{eq:energy-condition} can be neglected. At small and negative detuning ${\expos-\upsilon}$:
\begin{equation}
E \simeq -\frac{\hbar^2}{2\mref\Rstars\astars}.
\label{eq:E_s>1}
\end{equation}
This last law is almost reached for values of {the scaling exponent} $\expos$ of the order of ${1.4}$. This is illustrated in Fig.~(\ref{fig:spectrum_expos}), where the universal spectrum is plotted for increasing values of ${\expos>1}$.
For a small positive value of the detuning ${\expos-\upsilon}$, the generalized scattering length is large and negative in the vicinity of unitarity, giving a long-lived quasi-bound state when ${\expos \ge 1}$ \cite{large_v0p}. This corresponds to a complex energy solution of Eq.~\eqref{eq:energy-condition}: ${E=E_{\rm r}-{i\Gamma}/{2}}$. In the low energy limit, the resonance energy ${E_{\rm r}}$ is {still} given by Eq.~\eqref{eq:E_s>1}. {Moreover in the vicinity of the unitary limit where ${|\tau| \ll 1}$, one has ${r_{\ell,\expos,\tau} \simeq r_{\ell,\expos,0}}$ which is given by Eq.~\eqref{eq:r_ls0} and}
\begin{equation}
\Gamma \simeq  2 E_{\rm r}\left(\frac{2 \mref \Radius^2 E_{\rm r}}{\hbar^2} \right)^{\expos-1}  \frac{\Radius}{\Rstars} .
\label{eq:QB}
\end{equation}
{This last law is equivalent to that found in Refs.~\cite{Son22,Pri23,Pri24}.}
\section{Example of universality in a finite range model}

\subsection{Separable model potential}

The universality is now illustrated by using a reference model in a 2FI system with finite range interactions. The interaction ${V_{i3}}$ between the fermions ${i=1,2}$ and the impurity is separable:
{
\begin{equation}
V_{i3}= g  |\chi\rangle \langle \chi| 
\label{eq:V_separable}
\end{equation}
with the form factor
\begin{equation}
\langle \boldsymbol \kappa_1^{(i3)} |\chi\rangle= \left[1+\alpha (\kappa_{1}^{(i3)}\epsilon)^2\right]  \exp[-(\kappa_{1}^{(i3)}\epsilon)^2/2] 
\label{eq:form-factor}
\end{equation}
whereas the two identical fermions do not interact. The length ${\epsilon}$ gives the order of magnitude of the high energy scale ${\hbar^2/(\mref \epsilon^2)}$. From Eq.~\eqref{eq:Aref}, one has
\begin{equation}
\langle \boldsymbol \kappa_2^{(23)} | A^{\rm ref}  \rangle= \frac{\mref g}{2\pi \hbar^2} 
\int \frac{d^3 \kappa^{(23)}_1}{(2\pi)^3}
 \langle \chi | \boldsymbol \kappa^{(23)}_1 \rangle  \langle \boldsymbol \kappa^{(23)}_1,\boldsymbol \kappa_2^{(23)}| \Psi^{\rm ref} \rangle.
\label{eq:Aref_bis}
\end{equation}
In case of bound states, the function ${\langle \mathbf k |A^{\rm ref}_{23} \rangle}$ satisfies the STM-like equation
\begin{multline}
\frac{4\pi}{ \cos \theta} \int \frac{d^3k'}{(2\pi)^3} \frac{ \langle \chi|  \frac{\mathbf k'  + \mathbf k \sin \theta}{\cos \theta}  
\rangle \langle \frac{\mathbf k' \sin \theta + \mathbf k}{ \cos \theta}| \chi\rangle \langle \mathbf k'|A^{\rm ref}_{23}\rangle}{k'^2+k^2+2 \mathbf k \cdot \mathbf k' \sin \theta+ q^2 \cos^2 \theta}  \\
+ \langle \mathbf k|A^{\rm ref}_{23} \rangle \left[ -\frac{1}{\asc} + D(\sqrt{k^2+q^2})\right]
=0
 \label{eq:STM_Aref}
\end{multline}
where 
\begin{align}
&\chi(k)\equiv \langle \mathbf k |\chi\rangle= \left(1+ \alpha \epsilon^2 k^2 \right) \exp(-\epsilon^2 k^2/2) , \\
&D(k)= k \operatornamewithlimits{erfcx}( k \epsilon)+2 \alpha k^2 \epsilon \left[ \frac{1}{\sqrt{\pi}} -k \epsilon\operatornamewithlimits{erfcx}(k \epsilon)\right]\nonumber\\
&\qquad + \alpha^2 k^4 \epsilon^3  \left[ k\epsilon\operatornamewithlimits{erfcx}(k  \epsilon)
-\frac{1}{\sqrt{\pi}}
+\frac{1}{2k^2\epsilon^2\sqrt{\pi}}\right] 
\end{align}
and ${\operatornamewithlimits{erfcx}(u)=e^{u^2} \operatornamewithlimits{erfc}(u)}$ is the scaled complementary error function. Equation~\eqref{eq:STM_Aref} differs from the STM equation only for wavenumbers $k$ of the order of, or larger than ${1/\epsilon}$. In what follows, one considers the p-wave symmetry sector where ${\ell=1}$ and ${\langle \mathbf k| A^{\rm ref}_{23} \rangle=(\hat{\mathbf e}_{\mathbf k}\cdot\hat{\mathbf e}_{z}) \langle k | \mathcal A_1^{\rm ref} \rangle}$. More details about this finite range model and the resulting STM-like equation are given in Appendix \ref{app:ref-model}.}

{
 \subsection{ITBR threshold}
For a fixed value of the interaction parameter $\alpha$, it appears that for a sufficiently large mass ratio ${M/m \ge (M/m)_{\rm crit}<x^{\rm E}}$, the slight difference in the high-momentum limit, between the STM-like equation and the STM equation is able to induce a p-wave ITBR at unitarity. The critical mass ratio ${(M/m)_{\rm crit}}$ at which the ITBR occurs (which corresponds to the zero-energy limit of the resonant trimer at unitarity) is plotted as a function of the parameter ${\alpha}$ in Fig.~\ref{fig:ratio_alpha}.
\begin{figure}[h]
\centering
\includegraphics[width=8cm]{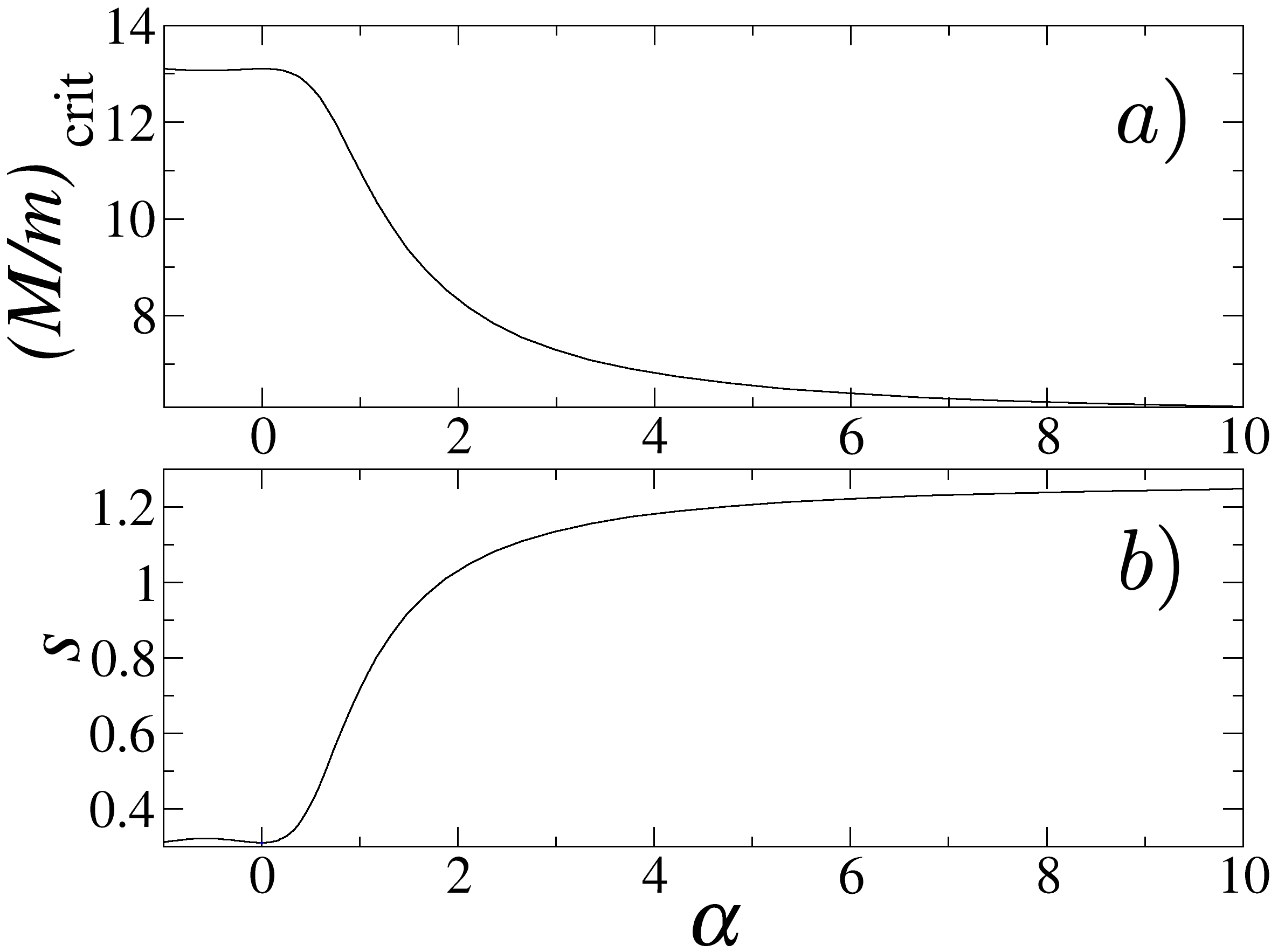}
\caption{{a) Critical mass ratio at which the ITBR occurs at unitarity  in the p-wave sector ${\ell=1}$ for the separable model of Eqs.~(\ref{eq:V_separable},\ref{eq:form-factor}), as a function of the parameter ${\alpha}$; b) same but 
for the value of the scaling exponent ${\expos}$.}}
\label{fig:ratio_alpha}
\end{figure}
It is interesting to notice that ITBR were found in the same configuration for a Gaussian or a P\"{o}schl-Teller potential in Ref.~\cite{Gan18}. For these potentials the critical mass ratio $(M/m)_{\rm crit}$ is of the order of $13$, i.e. a value close to the Efimov threshold ${x^{\rm E}}$ and to the one obtained with the separable potential used in this manuscript when ${\alpha=0}$.}
 
\begin{figure}
\centering
\includegraphics[width=8cm]{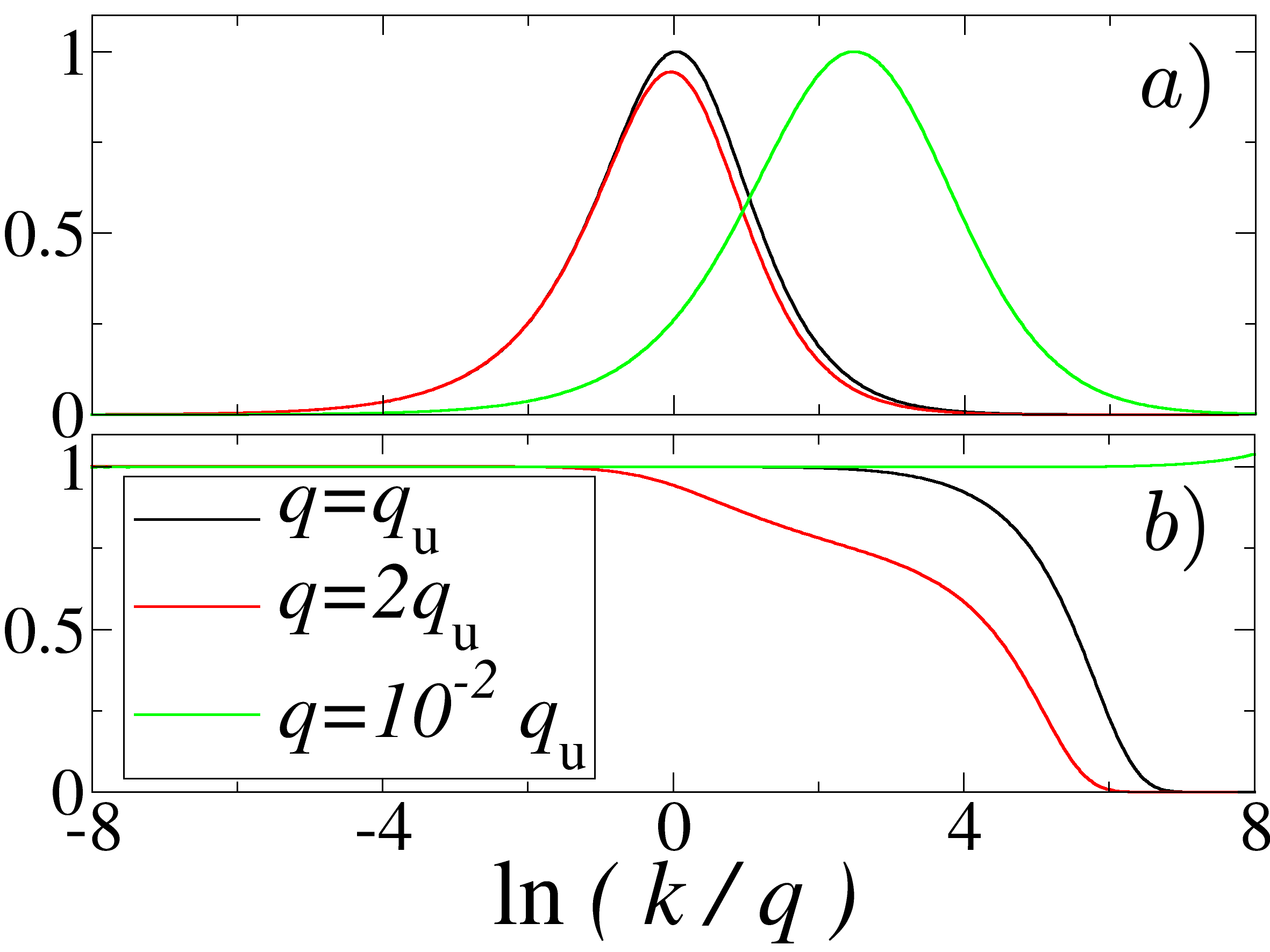}
\caption{(a) Plot of ${\langle k | \mathcal A_1^{\rm ref} \rangle}$ for three values of the binding wavenumber associated with the spectrum of Fig.~\ref{fig:1}  where ${\expos=0.25}$. The functions ${\langle k | \mathcal A_1 \rangle}$ for ${q=q_{\rm u}}$ and ${q=2q_u}$ coincide with the black solid line; (b) Plot of the ratio ${\langle k | \mathcal A_1^{\rm ref} \rangle/\langle k|\mathcal A_1 \rangle}$ for the same values of the binding wavenumber.
}
\label{fig:Aref_vs_A}
 \end{figure}

 {
 \subsection{Pure Gaussian separable potential}
 \label{sec:example_ITBR_0}
 One considers in this section the purely Gaussian separable potential corresponding to the case ${\alpha=0}$
 in Eq.~\eqref{eq:form-factor}. One finds an ITBR  for ${\ell=1}$ when ${\expos<0.3106\dots}$ corresponding to a mass ratio ${M/m \ge 13.10418\dots}$}. For such a small values of the scaling exponent, the centrifugal barrier in the hyperradius {is weak and is easily broken at short distance} by the attractive character of the {two-body} separable potential {which is in a s-wave resonant regime}. In the following, {to provide a tangible example}, one considers the case ${\expos=0.25}$ {corresponding to the mass ratio ${M/m=13.28\dots}$} where ${q_{\rm u} = 1.34\dots\times10^{-3}/\epsilon}$. The universality of the ${l=1}$ component, ${\langle k | \mathcal A_1^{\rm ref} \rangle}$ is illustrated in Fig.~\ref{fig:Aref_vs_A}. For ${q=q_u}$,  and ${q=10^{-2}\times q_u}$, with the proper normalization, ${\langle k | \mathcal A_1^{\rm ref} \rangle}$ is almost indistinguishable from ${\langle k | \mathcal A_1 \rangle}$ and the deviation from the universality is revealed by the plot of the ratio of these two functions. {For ${q=2q_u}$, the deviation from the universal function is more pronounced in the tail for ${k\gtrsim q}$.} 
{For a given ${\Radius}$, the parameters ${\upsilon}$ and ${\upsilon'}$ are identified by imposing that at unitarity, the spectrum of the contact model is tangent to the spectrum of the reference model. For such a small value of the scaling exponent ${\expos}$, calculations on the log-derivative condition Eq.~\eqref{eq:log-derivative} can be considered up to the order ${q^{2\expos}}$ and the second order terms in ${q^2}$ can be neglected. The choice of the radius $\Radius$ such that ${\upsilon'=0}$ appears then as the most natural one in the setting of the parameters. This is effectively what is observed in numerical calculations where ${\upsilon'=0}$ leads to the most accurate fit with the two other parameters given by ${\upsilon-\expos \simeq 6 \times 10^{-2}}$ and  ${\Radius/\epsilon \simeq 9.47}$. The corresponding effective range parameter is given by ${\Rstars/\epsilon=-15.8\dots}$ and at unitarity ${\epsilon/\astars \simeq 1.68 \times 10^{-2}}$. The spectrum is given by the green solid line in Fig.~\ref{fig:1}.
As expected, in this example the effective range term is not relevant: the approximation of  the universal spectrum obtained from Eq.~\eqref{eq:energy-condition} where $\Rstars$ is neglected and $\astars$ is given by Eq.~\eqref{eq:approx_as} is not distinguishable from the exact calculation at the scale of Fig.~\ref{fig:1}. The endpoint for the universal spectrum is located at ${\Radius/a^- \simeq -1.56\times 10^{-3}}$, which corresponds well to the value of  the reference model ${\Radius/a^- \simeq -1.57\times 10^{-3}}$. However, the approximate expression  in Eq.~\eqref{eq:endpoint_approx} is less precise giving ${\Radius/a^- \simeq -1.25\times 10^{-3}}$. In this regime where the scaling exponent $\expos$ is small and the parameter $\Rstars$ can be neglected, the spectrum obtained at finite two-body scattering length, generalizes the one-parameter law already found at unitarity in Refs.~\cite{Wer06b,Wer08,Gao15}.
}

{
\subsection{Example of spectrum for ${\expos>1}$}
\label{sec:s>1}
The same analysis as above can be performed for a value of the scaling exponent $\expos$ larger than unity. For instance, Fig.~\ref{fig:spectrum_s>1} displays a spectrum obtained for  the scaling exponent ${\expos=1.25}$ corresponding to the mass ratio ${M/m= 6.093\dots }$ with the interaction parameter just above the critical value of the ITBR threshold  ${\alpha_{\rm c}= 10.657517\dots}$ and ${\alpha=\alpha_c+10^{-5}}$. At unitarity, the binding wavenumber is ${q_u \epsilon=1.39\dots\times 10^{-4}}$. The optimal fit with the reference spectrum is obtained again for ${\upsilon'=0}$ with the detuning parameter
 ${\upsilon-\expos\simeq1.79  \times10^{-5}}$ and ${\Radius/\epsilon\simeq 22}$. The range parameter is given by ${\Rstars/\epsilon\simeq32.53 }$ and at the unitary limit, ${\epsilon/\astars \simeq 6.02  \times 10^{-7}}$. With these parameters, the position of the endpoint of the reference spectrum at
 ${\Radius/a^{-} \simeq -1.19 \times 10^{-5}}$ is very well approximated by Eq.~\eqref{eq:endpoint_approx}  which gives ${\Radius/a^{-} \simeq -1.20  \times 10^{-5}}$. In this example where $\expos$ is rather small, the spectrum is already in relative good agreement with the asymptotic law of Eq.~\eqref{eq:E_s>1} which is valid for large $\expos$ and  can be  expressed in terms of the binding wavenumber : 
\begin{equation}
q =\frac{1}{\sqrt{\astars \Rstars}} .
\label{eq:approx_spectrum_s>1}
\end{equation} 
Moreover, the approximations in Eqs.~(\ref{eq:approx_as},\ref{eq:approx_Rs}) have been used in the plot of the approximate spectrum in Fig.~\ref{fig:spectrum_s>1}. The functions ${\langle k | \mathcal A_1^{\rm ref} \rangle}$ are again compared to the functions ${\langle k | \mathcal A_1 \rangle}$ for a large range of values of $k$ in Fig.~\ref{fig:Aref_vs_A_1v25} and at the binding wavenumbers  ${q=q_u}$,  ${q=2q_u}$ and ${q=10^{-2}\times q_u}$. The detuning parameter is very small, explaining why for a given value of the binding wavenumber the distinction between the two functions is not visible in Fig.~\ref{fig:Aref_vs_A_1v25})-a. Plotting the ratio between the two functions in Fig.~\ref{fig:Aref_vs_A_1v25})-b makes it possible to exhibit a difference in the large momentum tail. 
\begin{figure}
\centering
\includegraphics[width=8cm]{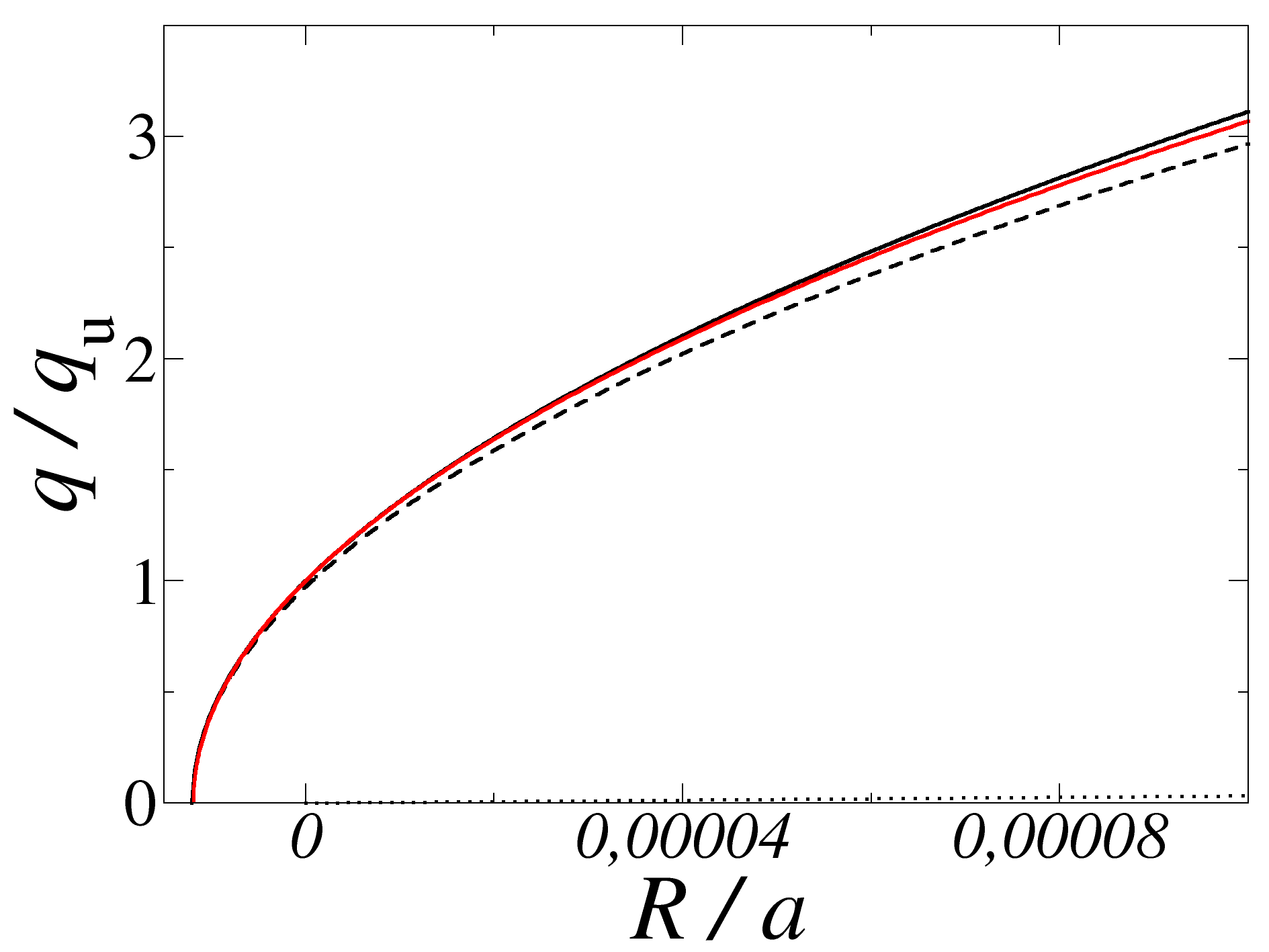}
\caption{{Plot of the ITBR spectrum (red line) of the reference model in Eqs.(\ref{eq:form-factor},\ref{eq:V_separable})  near the ITBR threshold for ${\expos = 1.25}$; Black line: universal spectrum 
%where the parameters $\upsilon$ and $\Radius$ have been adjusted in such a way that the two spectrum and their slope coincide at unitarity and ${\upsilon'=0}$. 
; Dash line: approximate spectrum of Eq.~\eqref{eq:approx_spectrum_s>1}; Dotted line: dimer spectrum. The numerical values of the parameters are given in the text of sec. \ref{sec:s>1}}}
\label{fig:spectrum_s>1}
 \end{figure}
} 
\begin{figure}
\centering
\includegraphics[width=8cm]{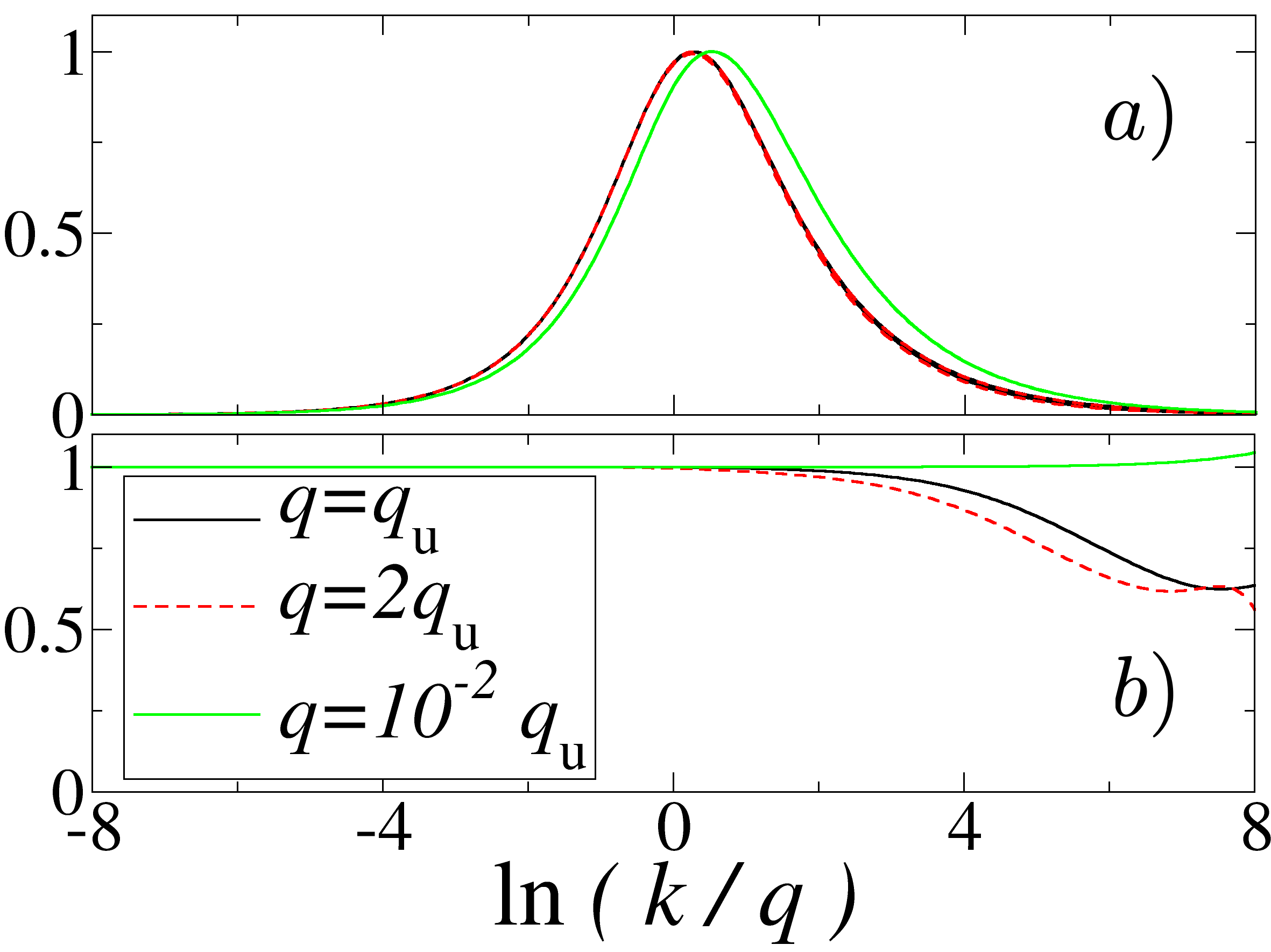}
\caption{{(a) Plot of ${\langle k | \mathcal A_1^{\rm ref} \rangle}$ for three values of the binding wavenumber associated with the spectrum in Fig.~\ref{fig:spectrum_s>1} where ${\expos=1.25}$. For each value of the binding wavenumber, the functions ${\langle k | \mathcal A_1 \rangle}$ are not distinguishable from ${\langle k | \mathcal A_1^{\rm ref} \rangle}$; (b) Plot of the ratio ${\langle k | \mathcal A_1^{\rm ref} \rangle/\langle k|\mathcal A_1 \rangle}$.}
}
\label{fig:Aref_vs_A_1v25}
 \end{figure}

\begin{figure}
\centering
\includegraphics[width=8cm]{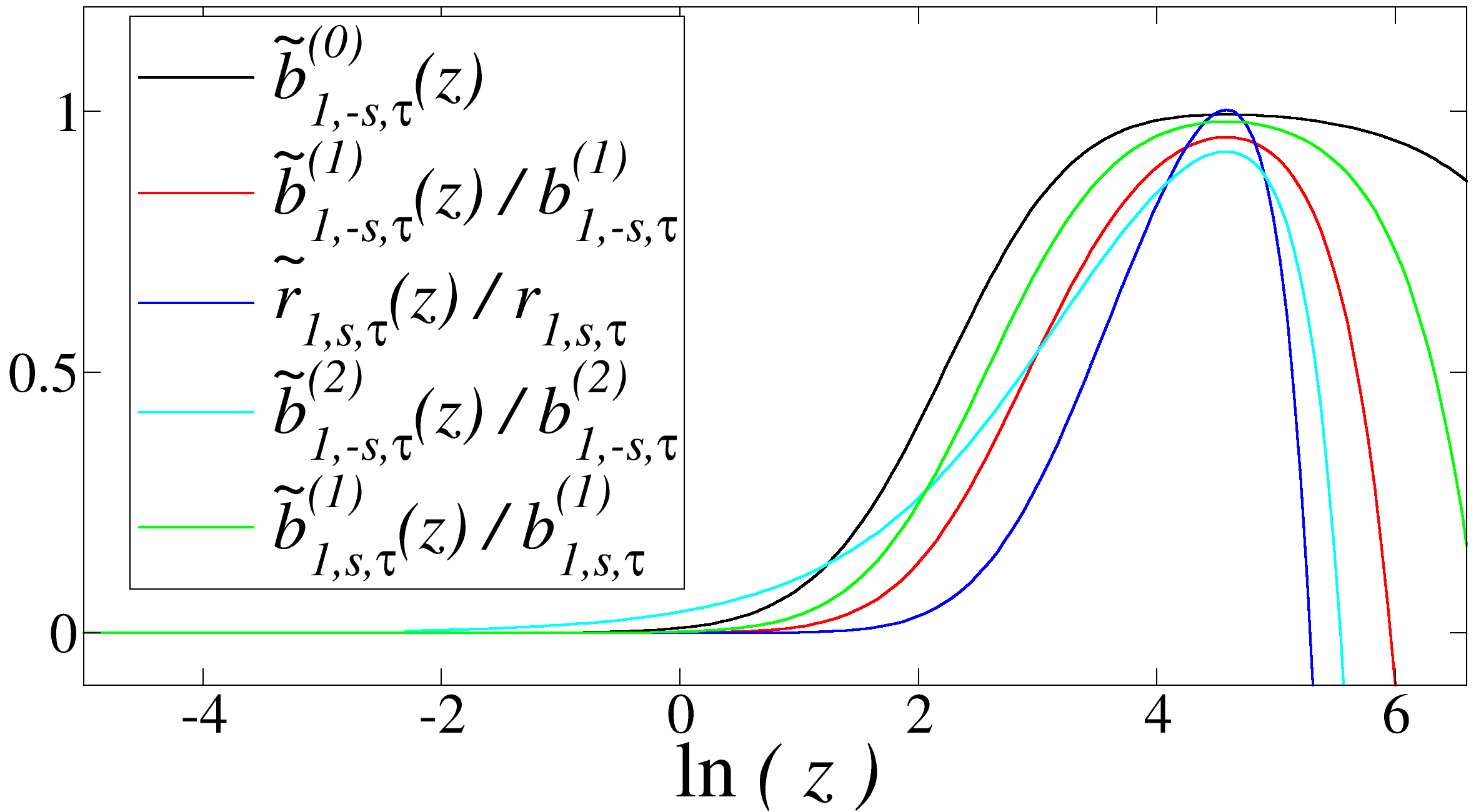}
\caption{{Plot of the ratio of the functions ${\tilde{b}_{1,\gamma,\tau}^{(n)}(z)}$ and ${\tilde{r}_{1,\expos,\tau}(z)}$ in Eq.~\eqref{eq:ansatz_Aref} with respect to their asymptotic value in the contact model with ${q=10^{-2}q_{\rm u}}$ and ${\expos=0.25}$. The parameters of the reference model correspond to those that give the spectrum near the endpoint  in Fig.~\ref{fig:spectrum_expos}.}}
\label{fig:plateaus_0v25}
\end{figure}

\begin{figure}
\centering
\includegraphics[width=8cm]{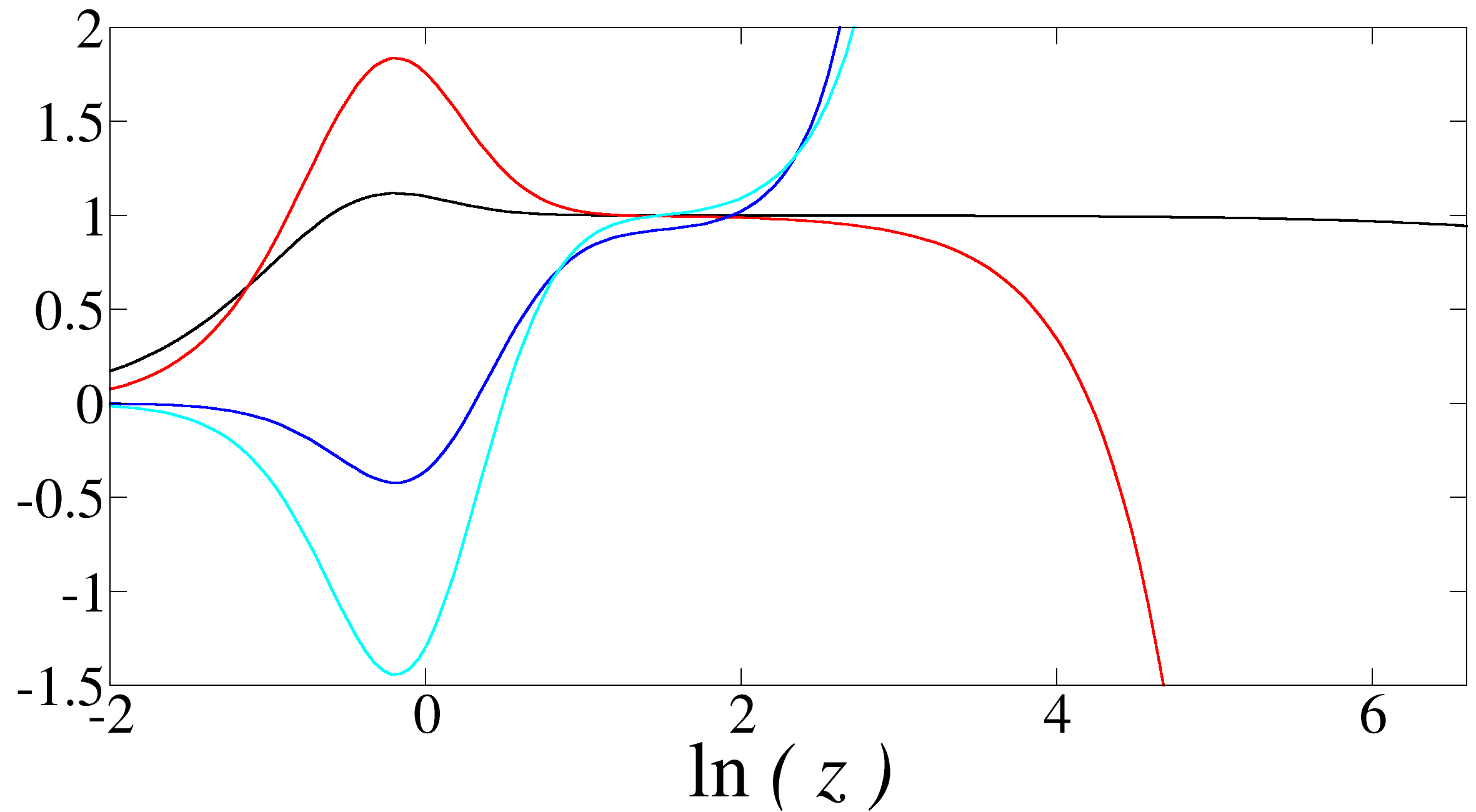}
\caption{{Same as Fig.~\ref{fig:plateaus_0v25} but for ${\expos=1.25}$. The parameters of the reference model correspond to those that give the spectrum near the endpoint in Fig.~\ref{fig:spectrum_s>1}}.}
\label{fig:plateaus_1v25}
\end{figure}

{
\subsection{Explaining why ${\upsilon'=0}$ is the best choice in this reference model}
\label{sec:best-choice}
If one considers the log-derivative of the reference function at the hyperradius $\Rsep$ including a linear energy correction:
\begin{equation}
\frac{\partial_\rho[\rho \langle \rho | \mathcal A_\ell^{\rm ref} \rangle]}{\langle \rho | \mathcal A_\ell^{\rm ref} \rangle} \biggr|_{\rho=\Rsep} =-\upsilon_0+\upsilon_0' q^2 \Rsep^2,
\label{eq:log-derivative-ref}
\end{equation}
then assuming that ${q\Rsep\ll 1}$, the parameter ${\upsilon_0'}$ is directly linked with the probability occupation at short distance \cite{Pri24}:
\begin{equation}
\int_{\rho<\Rsep} d^6\rho \, \left|\langle \boldsymbol \rho |\Psi^{\rm ref}(E) \rangle\right|^2 
= \upsilon'_0 \left| \Rsep^2 \langle \rho=\Rsep | \mathcal A_\ell^{\rm ref} \rangle \right|^2 .
\label{eq:sh_norm}
\end{equation}
Therefore, one can expect that a good choice for the log-derivative condition in Eq.~\eqref{eq:log-derivative} is ${\upsilon'=\upsilon_0'}$ and ${\Radius=\Rsep}$ where the radius $\Rsep$ is sufficiently large to have  ${\langle \rho | \mathcal A_\ell^{\rm ref} \rangle \simeq \langle \rho  | \mathcal A_\ell\rangle}$ when $\rho$ is in the vicinity $\Rsep$ (and with the  choice of normalization in Eq.~\eqref{eq:Psi_separable}). However, this identification has a sense only if the first terms of the expansion in Eq.~\eqref{eq:expansion_A} (at least up to ${n=2}$ in the first sum of the right hand side of this equation) lead to a good approximation of the reference function. To test this issue, let us compare the high momentum behavior of ${\langle k | \mathcal A_\ell^{\rm ref} \rangle}$ with the expansion of ${\langle k | \mathcal A_\ell \rangle}$ in the limit ${z=k/q \gg 1}$:
\begin{multline}
\langle k |A_\ell \rangle = \sum_{n=0}^\infty b_{\ell,-\expos,\tau}^{(n)} z^{\expos-2-n}
 -   \frac{\Gamma(1+\ell+\expos)r_{\ell,\expos,\tau}}{\Gamma(1+\ell-\expos)}\\
 \times  \sum_{n=0}^\infty b_{\ell,\expos,\tau}^{(n)} z^{-\expos-2-n} .
\label{eq:series_Ak}
\end{multline}
The coefficient ${b_{\ell,\gamma,\tau}^{(n)}}$ is proportional to ${c_{\ell,\gamma,\tau}^{(n)}}$ with a contant obtained from Eq.~\ref{eq:Hankel_power_law_appendix}. Concerning the reference function, one then uses the ansatz:
\begin{multline}
\langle k | \mathcal A_\ell^{\rm ref} \rangle= z^{-2+\expos} 
\left[ \tilde{b}_{\ell,-\expos,\tau}^{(0)}(z) +\frac{\tilde{b}_{\ell,-\expos,\tau}^{(1)}(z)}{z}+\frac{\tilde{b}_{\ell,-\expos,\tau}^{(2)}(z)}{z^2} \right]\\ 
+ \frac{\Gamma(1+\ell+\expos)\tilde{r}_{1,\expos,\tau}(z)}{\Gamma(1+\ell-\expos)}  z^{-2-\expos} \left[1+\frac{\tilde{b}_{\ell,\expos,\tau}^{(1)}(z)}{z} \right] .
\label{eq:ansatz_Aref}
\end{multline}
In the vicinity of a given value of ${z=k/q}$, the functions ${[\tilde{b}^{(n)}_{\ell,\expos,\tau}(z),\tilde{r}_{\ell,\expos,\tau}(z)]}$ are obtained by considering them as constant coefficients ({with respect to $z$}) and by solving a ${5 \times 5}$ system obtained from Eq.~\eqref{eq:ansatz_Aref} and its first four derivatives. In the region ${1 \ll z \ll 1/(q\epsilon)}$, one expects that the coefficients at the plateaus coincide with those of the universal function ${[{b}_{\ell,\gamma,\tau}^{(n)},{r}_{\ell,\expos,\tau}]}$. In what follows, using the reference model of this section, two examples have been considered for ${\ell=1}$  near the end point (at the binding wavenumber ${q =10^{-2} q_u}$) which is the most favorable situation to recover an equivalent behavior for the functions ${\langle k | \mathcal A_1^{\rm ref} \rangle}$ and ${\langle k | \mathcal A_1 \rangle}$ in the limit ${k/q \gg 1}$ while ${k \epsilon \ll 1}$. The degree of equivalence of the two functions in this limit has been illustrated by plotting the ratio ${\tilde{b}_{1,\pm\expos,\tau}(z)/b_{1,\pm\expos,\tau}}$ in Fig.~\ref{fig:plateaus_0v25} for ${\expos=0.25}$ and in Fig.~\ref{fig:plateaus_1v25} for ${\expos=1.25}$. In both figures, one observes a large plateau at large ${z}$ for ${\tilde{b}^{(0)}_{1,\expos,\tau}(z)}$ corresponding to the dominant ${z^{-2+\expos}}$ scaling behavior. However, due to finite range effects, there is no well defined plateau for the subdominant term in ${\tilde{b}^{(2)}_{1,\expos,\tau}(z)}$ at large momentum. However, the coefficient ${{b}^{(2)}_{1,\expos,\tau}}$ in the expansion of the universal function for a short hyperradius, gives the dependence of the log-derivative on the energy. Therefore, the absence of plateau for ${\tilde{b}^{(2)}_{1,\expos,\tau}(z)}$, in this standard situation of small detuning and low energy regime, shows that it is not possible to find a value of ${\Radius=\Rsep}$ such that ${\langle \rho | \mathcal A_\ell^{\rm ref} \rangle \simeq \langle \rho | \mathcal A_\ell \rangle}$ for ${q \Rsep  \sim \rho q \ll 1}$ and ${\upsilon_0'=\upsilon'}$. In other words, the equivalence of the two functions is not enough precise  to have an equivalence between the energy dependent log-derivatives considered at a value of ${\Rsep}$ satisfying ${q \Rsep \ll 1}$. When the energy of the state is varied by a small change in the inverse two-body scattering length ${1/\asc}$ near the unitarity, the short distance behavior of ${\langle \rho | \mathcal A_1^{\rm ref} \rangle}$ is marginally affected and thus from Eq.~\eqref{eq:sh_norm}, ${\upsilon_0'}$ remains almost fixed. However, there is no reason why the parameter ${\upsilon'}$ should remain fixed in this operation.}

{The optimal choice ${\upsilon'=0}$ found in numerical calculations follows from the fact that for this particular value of  ${\upsilon'}$, the parameter $\Radius$ is also tightly linked to the probability occupation at small distance through the identity \cite{Pri23}:
\begin{equation}
\int d^6\rho \, \left|\langle \boldsymbol \rho |\Psi^{\rm ref} \rangle\right|^2 
= \int_{\Radius<\rho} d^6\rho \, \left|\langle \boldsymbol \rho |\Psi \rangle\right|^2  
\label{eq:equiv_norm}
\end{equation}
where the tail of ${\langle \boldsymbol \rho |\Psi \rangle}$ is fixed by 
\begin{equation}
 \langle \boldsymbol \rho |\Psi \rangle \simeq \langle \boldsymbol \rho |\Psi^{\rm ref} \rangle
\qquad \text{for} \quad \rho \ge \Rsep .
\label{eq:tail}
\end{equation}
One can again argue that the short hyperradius behavior of ${\langle \rho | \Psi^{\rm ref} \rangle}$  remains almost fixed in a small change of the inverse scattering length ${1/\asc}$ near the unitarity and from Eq.~\eqref{eq:equiv_norm}, one expects the same property for the parameter ${\Radius}$. This is indeed what is observed in numerical calculations with the reference model: the value ${\upsilon'=0}$ leads to the best fit of the two spectrum for fixed and appropriate values of the parameters $\Radius$ and $\upsilon$, in a rather large interval of values of ${1/\asc}$ (and in particular near the endpoint at ${\asc=\asc^{-}}$).}

{One can notice that when ${\expos<1}$, the wavefunction of the contact model is square integrable and it is possible to encounter a situation where Eq.~\eqref{eq:equiv_norm} is not satisfied. This means that in the reference model the occupation probability at short hyperradius is anomalously large. This anomalous regime is analogous to a narrow s-wave resonance in the two-body problem \cite{Pri24}. In this case, the dependence of the log-derivative on the energy cannot be neglected and have to be taken into account to recover a good fit between the reference and the universal spectrum. Following the same reasoning as above, one expects that the optimal value for the triplet $(v,v',\Radius)$ is such that the normalization of the contact wavefunction obtained with the modified scalar product introduced in Ref.~\cite{Pri24} coincides with the normalization of the reference state:
\begin{equation}
\int d^6\rho \, \left|\langle \boldsymbol \rho |\Psi^{\rm ref} \rangle\right|^2 
= \int_{\Radius<\rho} d^6\rho \, \left|\langle \boldsymbol \rho |\Psi \rangle\right|^2  + v'  |\Radius^2 \mathcal A_\ell(\Radius)|^2
\label{eq:equiv_norm2}
\end{equation}
where Eq.~\eqref{eq:tail} holds. A detailed analysis of this regime must be carried out, accompanied by an example of a reference model giving rise to an ITBR in the anomalous regime, and is deferred to a future study.
}

\section{Conclusions}

With regard to the possible observation of ITBR in ultracold atoms, by using for instance the rf-association techniques  \cite{Mac12,Yud24}, the 2FI system composed of two ${^{171}}$Yb atoms and one caesium atom is an interesting candidate~\cite{Pri23}. As predicted generically in Ref.~\cite{Nai22}, an ITBR should occur for ${\ell=1}$, from the combination of an Yb-Yb optical p wave Feshbach resonance and a Yb-Cs magnetic s-wave Feshbach resonance \cite{Goy10,Yama13,Yang19}. Using other species offers the possibility of exploring the spectrum for other mass ratios~\cite{Zhao24}. Recently another analysis using  Lennard-Jones potentials confirmed  the existence of ITBR in 2FI systems with p-wave interactions~\cite{Chen25}. In these examples, the two-body resonance results from a short-range attraction between the two identical fermions that allows the two-body centrifugal barrier characteristic of the p-wave to be overcome at small distance. In the three-body system, qualitatively speaking, it is this same two-body attraction that again enables the centrifugal barrier to be overcome at small hyperradius, leading to a resonance. On this basis, a similar effect is expected for two-body resonances in higher partial waves. An alternative scenario to achieve an ITBR is given by a Feshbach coupling between the three particles in an open channel and a three-body molecular state in a closed channel. %This work paves the way to find universal laws at finite scattering length for isolated $N$-body resonances in other systems.

\section*{Acknowledgments}

I thank Yvan Castin, Pascal Naidon and Félix Werner for remarks and discussions.

\ 
\newpage

\begin{widetext}

\appendix

%These appendices  gathers some of the technical details and steps used in deriving the results from the main text. 

\section{Jacobi coordinates}
\label{app:Jacobi}

The Jacobi coordinates used in this work are a particular choice of their general definition which are remained in what follows. For a given reference mass $\mref$, the three systems of Jacobi coordinates associated with each possible pair of particles $(ij)$ are defined as follows
\begin{equation}
\boldsymbol \rho^{(ij)}_1 = \sqrt{\frac{\mu_{ij}}{\mref}}  (\mathbf r_j - \mathbf r_i) 
\ ; \ 
\boldsymbol \rho^{(ij)}_2 = \sqrt{\frac{\mu_{ij,l}}{\mref}} \left( \mathbf r_l - \frac{m_i \mathbf r_i+ m_j \mathbf r_j}{m_i+m_j}\right)
\ ; \ 
\boldsymbol \rho^{(ij)}_3 =   \frac{m_1 \mathbf r_1+m_2 \mathbf r_2+m_3 \mathbf r_3}{\sqrt{\mref (m_1+m_2+m_3)}},
\label{eq:Jacobi_N=3}
\end{equation}
where the three integers ${(i,j,l)}$ represent cyclic permutations of the triplet ${(123)}$ and the reduced masses are given by
\begin{equation}
\mu_{ij} = \frac{m_im_j}{m_i+m_j}\quad  ; \quad \mu_{ij,l} = \frac{m_l(m_i+m_j)}{m_1+m_2+m_3} .
\end{equation}
In what follows, when the index $^{(ij)}$ for the Jacobi coordinates ${(\boldsymbol \rho_1,\boldsymbol \rho_2)}$ is not specified, this means that the relations remain valid whatever the choice of set of coordinates. For instance, the hyperradius in the center of mass frame $\rho=\sqrt{(\rho_1)^2+(\rho_2)^2}$ is invariant in the choice of the pair $(ij)$:
\begin{equation}
\rho^2=\sum_{i=1}^3 \frac{m_i}{\mref} ( \mathbf r_i-\mathbf C)^2,
\end{equation}
where ${\mathbf C}$ is the center of mass:
\begin{equation}
\mathbf C =\frac{ m_1 \mathbf r_1+ m_2\mathbf r_2 + m_3\mathbf r_3}{ m_1+ m_2+ m_3}.
\end{equation}
The three wavevectors which are the conjugate of the Jacobi coordinates are given by
\begin{equation}
\boldsymbol \kappa^{(ij)}_1 = \sqrt{\frac{\mref}{\mu_{ij}}}  \frac{m_i \mathbf k_j-m_j\mathbf k_i}{m_i+m_j} 
\ ; \ 
\boldsymbol \kappa^{(ij)}_2 = \sqrt{\frac{\mref}{\mu_{ij,l}}}  \frac{(m_i+m_j)\mathbf k_l - m_l(\mathbf k_i+\mathbf k_j)}{m_1+m_2+m_3}
\ ; \ 
\boldsymbol  \kappa^{(ij)}_3=\sqrt{\frac{\mref}{m_1+m_2+m_3}}\left( \mathbf k_1+ \mathbf k_2+ \mathbf k_3\right) .
\end{equation}
For a given set, by construction, they are such that 
\begin{equation}
\sum_{n=1}^3 \boldsymbol  \kappa_n \cdot  \boldsymbol  \rho_n = \sum_{n=1}^3  \mathbf k_n \cdot  \mathbf r_n .
\end{equation}
In the center of mass frame, the hyperwavenumber ${\kappa=\sqrt{\kappa_1^2+\kappa_2^2}}$ satisfies also
\begin{equation}
\kappa^2=\sum_{i=1}^3 \frac{\mref}{m_i} k_i^2 .
\end{equation}
In this work, one considers mass imbalanced systems where ${m_1=m_2=M}$ and ${m_3=m}$ with the mass ratio ${x=M/m}$ and all the equations are derived with the choice of reference mass $\mref=\mu_{13}=\mu_{23}$ which simplifies all the expressions. In the center of mass frame, only the two vectors ${(\boldsymbol \rho_1^{(ij)},\boldsymbol \rho_2^{(ij)})}$ are relevant. {Introducing the relative coordinates ${\mathbf r_{ij}=\mathbf r_j - \mathbf r_i}$,} they are given by
\begin{equation}
\left\{
\begin{array}{l}
 \boldsymbol \rho^{(23)}_1 = \mathbf r_{23}\\
 \boldsymbol \rho^{(23)}_2 = \frac{1+x}{\sqrt{1+2x}}\left( \mathbf r_1 - \frac{x \mathbf r_2+ \mathbf r_3}{1+x}\right)
\end{array}
\right.
\ ; \ 
\left\{
\begin{array}{l}
\boldsymbol \rho^{(13)}_1 = \mathbf r_{13}  \\
\boldsymbol \rho^{(13)}_2 = \frac{1+x}{\sqrt{1+2x}}\left( \mathbf r_2 - \frac{x \mathbf r_1+ \mathbf r_3}{1+x}\right)
\end{array}
\right. 
\ ; \ 
\left\{
\begin{array}{l}
\boldsymbol \rho^{(12)}_1 = \sqrt{\frac{1+x}{2}}  \mathbf r_{12} \\
\boldsymbol \rho^{(12)}_2 = \sqrt{\frac{2(1+x)}{1+2x}}
\left( \mathbf r_3 - \frac{ \mathbf r_1+ \mathbf r_2}{2}\right)
\end{array}
\right.
\end{equation}
and the wavevectors of the three sets are given by
\begin{equation}
\left\{
\begin{array}{l}
\boldsymbol \kappa^{(23)}_1 = \frac{x \mathbf k_3-\mathbf k_2}{1+x}\\ 
 \boldsymbol \kappa^{(23)}_2 =  \frac{\sqrt{1+2x}}{1+x} \mathbf k_1
\end{array}
\right. 
\ ; \ 
\left\{
\begin{array}{l}
 \boldsymbol \kappa^{(13)}_1 = \frac{x \mathbf k_3-\mathbf k_1}{1+x} \\
 \boldsymbol \kappa^{(13)}_2 = \frac{\sqrt{1+2x}}{1+x} \mathbf k_2 
\end{array}
\right. 
\ ; \ 
\left\{
\begin{array}{l}
\boldsymbol \kappa^{(12)}_1 = \frac{\mathbf k_2-\mathbf k_1}{\sqrt{2(1+x)}} \\
\boldsymbol \kappa^{(12)}_2 = \sqrt{\frac{1+2x}{2(1+x)}} \mathbf k_3.
\end{array}
\right. 
\end{equation}
For convenience one also introduces the angle ${\theta \in [0,\pi/2]}$ which satisfies
\begin{equation}
\sin \theta = \frac{x}{1+x}\quad , \quad \cos \theta=\frac{\sqrt{1+2x}}{1+x} .
\end{equation}
With this notation, the Jacobi coordinates are defined by:
\begin{equation}
\boldsymbol \rho^{(23)}_2 = \mathbf r_{23} \tan \theta + \mathbf r_{31} \sec \theta  \quad ; \quad  \boldsymbol \rho^{(13)}_2 = \mathbf r_{13} \tan \theta + \mathbf r_{32} \sec \theta
\label{eq:rho_2}
\end{equation}
and
\begin{equation}
\boldsymbol \rho^{(12)}_1 = \mathbf r_{12} \cos\left(\frac{\pi}{4}-\frac{\theta}{2} \right) \sec \theta \quad \text{or} \quad
 \boldsymbol \rho^{(12)}_2 = \left( \mathbf r_{13} + \mathbf r_{23}\right)
\sin\left(\frac{\pi}{4}-\frac{\theta}{2} \right)  \sec \theta ,
\label{eq:rho_12}
\end{equation}
where one has used the two identities:
\begin{equation}
\cos\left(\frac{\pi}{4}-\frac{\theta}{2} \right) = \sqrt{\frac{1+\sin \theta}{2}} \quad ; \quad 
\sin\left(\frac{\pi}{4}-\frac{\theta}{2} \right) = \sqrt{\frac{1-\sin \theta}{2}} .
\end{equation}
Introducing the ${2\times2}$ active rotation matrix
\begin{equation}
\mathcal R(\alpha) =
\left[
\begin{array}{cc}
\cos \alpha & - \sin \alpha\\
\sin \alpha &  \cos \alpha
\end{array}
\right] ,
\end{equation}
the different set of coordinates are related the ones to the others by a rotation:
\begin{equation}
\left[
\begin{array}{c}
\boldsymbol \rho^{(12)}_1 \\
\boldsymbol \rho^{(12)}_2
\end{array}
\right]
=\mathcal R\left(\frac{3\pi}{4}-\frac{\theta}{2}\right)
\left[
\begin{array}{c}
\boldsymbol \rho^{(23)}_1 \\
\boldsymbol \rho^{(23)}_2
\end{array}
\right]
\ ; \ 
\left[
\begin{array}{c}
\boldsymbol \rho^{(13)}_1 \\
\boldsymbol \rho^{(13)}_2
\end{array}
\right]
=\mathcal R\left(\frac{\pi}{2}+\theta\right)
\left[
\begin{array}{c}
\boldsymbol \rho^{(32)}_1 \\
\boldsymbol \rho^{(32)}_2
\end{array}
\right] 
\ ; \ 
\left[
\begin{array}{c}
\boldsymbol \rho^{(13)}_1 \\
\boldsymbol \rho^{(13)}_2
\end{array}
\right]
=\mathcal R\left(\frac{\theta}{2}-\frac{3\pi}{4}\right)
\left[
\begin{array}{c}
\boldsymbol \rho^{(21)}_1 \\
\boldsymbol \rho^{(21)}_2
\end{array}
\right] .
\end{equation}
One has also the property
\begin{equation}
\left[
\begin{array}{c}
\boldsymbol \rho^{(ij)}_1 \\
\boldsymbol \rho^{(ij)}_2
\end{array}
\right]
=\mathcal R\left(\alpha \right)
\left[
\begin{array}{c}
\boldsymbol \rho^{(kl)}_1 \\
\boldsymbol \rho^{(kl)}_2
\end{array}
\right]
\quad \Longleftrightarrow \quad
\left[
\begin{array}{c}
\boldsymbol \rho^{(ji)}_1 \\
\boldsymbol \rho^{(ji)}_2
\end{array}
\right]
=\mathcal R\left(-\alpha \right)
\left[
\begin{array}{c}
\boldsymbol \rho^{(lk)}_1 \\
\boldsymbol \rho^{(lk)}_2
\end{array}
\right] .
\label{eq:transfos_rho}
\end{equation}
Identities analogous to Eqs.~(\ref{eq:rho_2}-\ref{eq:transfos_rho}) old for the hyperwavevectors ${\boldsymbol \kappa_n^{(ij)}}$.

\section{STM equation}
\label{app:STM}

Although the Skorniakov Ter-Martirosian (STM) equation is well known, a derivation is proposed in this Appendix. The stationary Schr\"{o}dinger equation in the configuration space for a state of energy ${E=-\frac{\hbar^2q^2} {2\mref}}$ is: 
\begin{equation}
\left(q^2 - \Delta_{\boldsymbol \rho_1} - \Delta_{\boldsymbol \rho_2}\right) \langle \boldsymbol \rho | \Psi \rangle  =-
 4 \pi \sum_{i<j} \delta(\boldsymbol \rho^{(ij)}_1) \langle \boldsymbol \rho^{(ij)}_2 |A_{ij}\rangle .
 \label{eq:Schrodi_rho}
\end{equation}
The delta distributions in the right-hand-side of Eq.~\eqref{eq:Schrodi_rho} are obtained from the action of a three dimensional Laplacian on the inverse radius singularity inherent of the Wigner-Bethe-Peierls contact condition of two interacting particles ${(ij)}$ when ${\rho^{(ij)}_1 \to 0}$, i.e.:
\begin{equation}
\langle \boldsymbol \rho | \Psi \rangle =  \langle \boldsymbol \rho^{(ij)}_2 | A_{ij} \rangle \left(\frac{1}{a}-\frac{1}{\rho^{(ij)}_1} \right) + O(\rho^{(ij)}_1) .
\label{eq:BP_ij}
\end{equation}
In the reciprocal space one uses the wavevectors ${(\boldsymbol \kappa^{(ij)}_1,\boldsymbol \kappa^{(ij)}_2)}$ which are the conjugate of ${(\boldsymbol \rho^{(ij)}_1,\boldsymbol \rho^{(ij)}_2)}$. For convenience, one uses the short-hand notations  ${(\mathbf p,\mathbf k)=(\boldsymbol \kappa^{(23)}_1,\boldsymbol \kappa^{(23)}_2)}$, ${\mathbf k'=\boldsymbol \kappa^{(13)}_2}$ and ${\mathbf k''=\boldsymbol \kappa^{(12)}_2}$. In the center of mass frame:
\begin{equation}
\mathbf p =-\mathbf k_1 \sin \theta -\mathbf k_2 \ ; \  \mathbf k= \mathbf k_1 \cos \theta .
\end{equation}
The Fourier transform of Eq.~\eqref{eq:Schrodi_rho} gives for negative energy states :
\begin{equation}
\langle \mathbf p,\mathbf k | \Psi \rangle
 =-\frac{4\pi}{q^2 + k^2 + p^2} \left( \langle \mathbf k | A_{23} \rangle + \langle \mathbf k' | A_{13} \rangle + \langle \mathbf k'' | A_{12} \rangle \right).
\label{eq:Schrodi_1}
\end{equation}
Considering the contact condition in Eq.~\eqref{eq:BP_ij} for the pair ${(23)}$, one remarks that the singularity ${1/\rho_1^{(23)}}$ is of the same type as the singularity of the two-body Green's function of the Helmholtz equation at the imaginary wavenumber ${i\lambda}$: 
\begin{equation}
\int \frac{d^3p}{(2\pi)^3} \frac{e^{i\mathbf p \cdot \boldsymbol \rho^{(23)}_1}}{p^2+\lambda^2} =
\frac{\exp(-\lambda \rho^{(23)}_1)}{4\pi \rho^{(23)}_1} .
\label{eq:Green}
\end{equation}
The contact condition, can be then implemented in the reciprocal space in few steps. In a first step one subtracts the inverse radius singularity of ${ \langle \boldsymbol \rho | \Psi \rangle}$ when ${\rho^{(23)}_1\to 0}$ and rewrites Eq.~\eqref{eq:BP_ij} as 
\begin{equation}
 \int \frac{d^3pd^3k}{(2\pi)^6} \left( e^{i\left(\mathbf p \cdot \boldsymbol \rho^{(23)}_1+\mathbf k \cdot \boldsymbol \rho^{(23)}_2\right)} \langle \mathbf p,\mathbf k | \Psi \rangle\right) + 4\pi   \langle \boldsymbol \rho^{(23)}_2 | A_{23} \rangle \int \frac{d^3p}{(2\pi)^3} \frac{e^{i\mathbf p \cdot \boldsymbol \rho^{(23)}_1}}{p^2+\lambda^2} = \langle \boldsymbol \rho^{(23)}_2 | A_{23} \rangle \left(\frac{1}{a} -\lambda \right) + O(\rho^{(23)}_1)
\label{eq:BP_bis}
 \end{equation}
where the parameter ${\lambda>0}$ is arbitrary. One then takes the limit ${\rho^{(23)}_1 \to 0}$ and performs a Fourier transform with respect to the variable $\boldsymbol \rho^{(23)}_2$ in this last equation:
\begin{equation}
\int \frac{d^3p}{(2\pi)^3} \left[\langle \mathbf p,\mathbf k | \Psi \rangle +  \frac{4\pi\langle \mathbf k| A_{23} \rangle }{p^2+\lambda^2} \right] 
=\langle \mathbf k | A_{23} \rangle\left(\frac{1}{\asc} -\lambda \right) ,
\label{eq:STM_lambda}
\end{equation}
Injecting Eq.~\eqref{eq:Schrodi_1} in the integral term of the left-hand-side of Eq.~\eqref{eq:STM_lambda}  with  the choice ${\lambda=\sqrt{k^2+q^2}}$ gives:
\begin{equation}
-4\pi \int \frac{d^3 p}{(2\pi)^3} 
\frac{ \langle  \mathbf k' | A_{13} \rangle +  \langle  \mathbf k'' | A_{12} \rangle }{p^2 +k^2 + q^2}\\
 =  \langle  \mathbf k  | A_{23} \rangle  \left(\frac{1}{\asc}- \sqrt{k^2+ q^2}\right) .
\label{eq:STM_3body}
\end{equation}
Depending on the system studied and using the exchange symmetry, the states ${|A_{ij}\rangle}$ are expressed in terms of the single state ${|A\rangle}$ as given in Tab.~(\ref{tab:systems}).
\begin{table}
\begin{tabular}{|c|c|c|c|}
\hline
System & Symmetry &\ $\stat$\ \ & ${{M}/{m}}$ \\
\hline
3B &$|A_{12}\rangle=|A_{13}\rangle= |A_{23}\rangle=|A \rangle$& 2 & x=1\\
\hline
2BI &$|A_{12}\rangle=0$ $|A_{23}\rangle= |A_{13}\rangle=|A \rangle$& 1 & x \\
\hline
2FI &$|A_{12}\rangle=0$ $|A_{23}\rangle=- |A_{13}\rangle=|A \rangle$& -1 &x \\
\hline
\end{tabular}
\label{tab:systems}
\caption{Three types of systems are considered in the STM equation: three bosons (3B), two bosons and one impurity (2BI), two fermions and one impurity (2FI). The masses of the three particles are given by ${m_1=m_2=M}$ and ${m_3=m}$.}
\end{table}
Equation \eqref{eq:STM_3body} can be then simplified as
\begin{equation}
4\pi \stat \int \frac{d^3 p}{(2\pi)^3} 
 \frac{  \langle  \mathbf k' | A \rangle}{p^2+k^2+q^2} =  \langle  \mathbf k  | A \rangle \left( \sqrt{k^2+ q^2}-\frac{1}{\asc}\right),
\label{eq:STM_3body_beta}
\end{equation}
where the statistical factor ${\stat}$ is also given in Tab.~(\ref{tab:systems}). The coordinates ${\mathbf p}$ are then expressed as a function of ${\mathbf k}$ and ${\mathbf k'}$:
\begin{equation}
{\mathbf p}  =-\mathbf k' \sec \theta -  \mathbf k \tan \theta \quad ,
\end{equation}
which gives 
\begin{equation}
d^3p = \frac{d^3k'}{\cos^3 \theta} \quad ; \quad p^2+k^2=\frac{k'\,^2+k^2+2 \mathbf k\cdot \mathbf k' \sin \theta}{\cos^2\theta}.
\label{eq:d3p}
\end{equation}
From Eqs.~(\ref{eq:STM_3body_beta},\ref{eq:d3p}), one obtains the STM equation:
\begin{equation}
 \int \frac{d^3 k'}{(2\pi \cos \theta)^3} 
\frac{ 4\pi \stat \langle \mathbf k' | A  \rangle}{q^2 +\left(k'\,^2+k^2+2 \mathbf k \cdot \mathbf k' \sin \theta \right) \sec^2 \theta }
 = \langle \mathbf k | A \rangle \left( \sqrt{k^2+q^2}-\frac{1}{\asc} \right) .
\label{eq:STM_bis}
\end{equation}
This last equation is invariant in a rotation of the coordinates. For a given value of the orbital momentum ${\ell \hbar}$, one defines the $\ell$-wave component of the function $\langle \mathbf k|A\rangle$: 
\begin{equation}
\langle \mathbf k| A  \rangle=P_\ell(\hat{\mathbf e}_{\mathbf k}\cdot\hat{\mathbf e}_{z}) \langle k | \mathcal A_\ell \rangle .
\end{equation}
Projection of Eq.~\eqref{eq:STM_bis} on the $\ell$-wave gives 
\begin{equation}
 \int_0^\infty \frac{dk'}{\cos \theta} \int_{-1}^1 \frac{du}{\pi} 
\frac{\stat  k'^2  \langle k'| \mathcal A_\ell \rangle
P_\ell(u)}{q^2 \cos^2 \theta + k'\,^2+k^2+2u kk' \sin \theta}
=\langle k | \mathcal A_\ell \rangle \left(\sqrt{k^2+q^2}-\frac{1}{a}\right) .
\label{eq:STM_3D_l}
\end{equation}
The last equation is rewritten in dimensionless form as 
\begin{equation}
 \int_0^\infty \frac{dz'}{\cos \theta} \int_{-1}^1 \frac{du}{\pi} 
\frac{\stat  z'\,^2 \langle z' | \phi_{\ell,\expos,\tau}\rangle P_\ell(u)}{\cos^2\theta + z'\,^2+z^2+2 u z z'  \sin \theta }
 = \langle z | \phi_{\ell,\expos,\tau}\rangle \left(\sqrt{z^2+1}-\tau\right) ,
\label{eq:STM_phi}
\end{equation}
where ${z=k/q}$, ${\tau=1/(q\asc)}$ and  ${\langle z | \phi_{\ell,\expos,\tau}\rangle}$ is a particular normalization of the function ${\langle k | \mathcal A_\ell \rangle }$. The angular integration can be expressed in terms of the function 
\begin{equation}
M_\ell(t)= \frac{1}{2} \int_{-1}^{1} du \frac{P_\ell(u)}{t+u} \quad ,
\end{equation}
where ${t=(z^2+z'\,^2+\cos^2 \theta)/(2zz')>1}$. One has
\begin{equation}
M_0(t)=\frac{1}{2} \ln \left( \frac{t+1}{t-1}\right)=Q_0(t)
\quad ; \quad 
M_1(t)=1-\frac{t}{2} \ln \left( \frac{t+1}{t-1}\right)=-Q_1(t) 
\label{eq:M01}
\end{equation}
and for ${l \ge 2}$, the function ${M_\ell}$ satisfies the recurrence relation 
\begin{equation}
M_\ell(t)=\frac{-(2\ell-1)t}{\ell} M_{\ell-1}(t) - \frac{(\ell-1)}{\ell}  M_{\ell-2}(t)  \quad ,
\label{eq:M_recurrence}
\end{equation}
which is the same as the one for ${(-1)^\ell Q_\ell(t)}$ and from the explicit expressions in Eqs.~\eqref{eq:M01}, one finds
\begin{equation}
M_\ell(t) =(-1)^\ell Q_\ell(t) .
\end{equation}
Finally, the STM equation can  be recasted  in the following form:
\begin{equation}
\frac{2\stat (-1)^\ell }{\pi \sin (2 \theta)} \int_0^\infty \frac{z'dz'}{z}  
  Q_\ell\left(\frac{z^2+z'\,^2+\cos^2\theta}{2 z z' \sin \theta } \right) \langle z' | \phi_{\ell,\expos,\tau}\rangle 
 = \langle z | \phi_{\ell,\expos,\tau}\rangle \left(\sqrt{z^2+1}-\tau\right) .
\label{eq:STM_compact}
\end{equation}
Introducing the operator
\begin{equation}
\langle z | \mathcal L^\Lambda| \phi_{\ell,\expos,\tau} \rangle =  \sqrt{z^2+1} \langle z | \phi_{\ell,\expos,\tau} \rangle
- \frac{2\stat (-1)^\ell }{\pi \sin (2 \theta)} \int_0^\Lambda \frac{z'dz'}{z}  
 Q_\ell\left(\frac{z^2+z'^2+\cos^2\theta}{2 z z' \sin \theta } \right) \langle z' | \phi_{\ell,\expos,\tau} \rangle ,
\label{eq:L_infty}
\end{equation}
one obtains the STM equation in the main text.

{
\section{Separable model}
\label{app:ref-model}
The two-body scattering amplitude of the interaction potential in Eqs.~(\ref{eq:V_separable},\ref{eq:form-factor}) is purely s-wave and at the relative energy ${E=\frac{\hbar^2k^2}{2 \mu_{23}}}$ in the center of mass of the impurity-fermion system, it is given by
\begin{equation}
f_{\rm s}(k) = -\frac{\chi(k)^2}{{1}/{\asc} +i k\, w(k\epsilon)+2 \alpha k^2 \epsilon \left[ ik \epsilon\, w(k\epsilon) + \frac{1}{\sqrt{\pi}}  \right] + \alpha^2 k^4 \epsilon^3  \left[ i k \epsilon\, w(k\epsilon)
+\frac{1}{\sqrt{\pi}} + \frac{1}{2k^2\epsilon^2\sqrt{\pi}}\right]}\ 
\end{equation}
where ${w(u)=\operatornamewithlimits{erfcx}(-iu)}$ is the Faddeeva function and the scattering length is given by 
\begin{equation}
\asc=\frac{1}{\frac{2\pi \hbar^2}{g\mu_{23}} + \frac{1}{\epsilon \sqrt{\pi} \left(1+\alpha +\frac{3\alpha^2}{4} \right)} }
\end{equation}
To derive the STM-like equation in Eq.~\eqref{eq:STM_Aref}, one  writes the stationary Schr\"{o}dinger equation of the reference model in the reciprocal space:
\begin{equation}
(\kappa^2 +q^2) \langle  \boldsymbol \kappa | \Psi^{\rm ref} \rangle
= -4\pi \langle \kappa^{(23)}_1 | \chi \rangle \langle \boldsymbol \kappa^{(23)}_2 | A^{\rm ref} \rangle 
+ 4\pi \langle \kappa^{(13)}_1 | \chi \rangle  \langle \boldsymbol \kappa^{(13)}_2 | A^{\rm ref} \rangle .
\label{eq:Schrodi_ref}
\end{equation}
and injects it in the definition of ${\langle \mathbf k | A^{\rm ref}  \rangle}$ given in Eq.~\eqref{eq:Aref_bis}. In the p-wave sector where  ${\langle \mathbf k | A^{\rm ref}  \rangle=(\hat{\mathbf e}_{\mathbf k}\cdot \hat{\mathbf e}_z) \langle k | \mathcal A_1^{\rm ref}  \rangle}$, the STM-like equation reads
\begin{equation}
\frac{|\chi(iq_{\rm rel})|^2 \langle k | \mathcal A_1^{\rm ref}  \rangle}{f_{\rm s}(iq_{\rm rel})} +\frac{2}{\pi}  \int_0^\infty \frac{k'dk'}{k} 
 \frac{e^{ -\epsilon^2 (k'^2+k^2) (\sec^2\theta-\frac{1}{2})}}{ \sin(2 \theta)}\\
  F\left(a_1,a_2,a_3,a_4 \right)   \langle k' | \mathcal A_1^{\rm ref}  \rangle=0
\label{eq:STM_ref_l=1}
\end{equation}
where ${q_{\rm rel}=\sqrt{k^2+q^2}}$ and the function ${F}$ is defined by
\begin{equation}
F(a_1,a_2,a_3,a_4)=\frac{1}{2} \int_{-1}^1dX \frac{\left(a_1X+a_2 X^2+a_3 X^3 \right) e^{-a_4 X}}{a_5+X} .
\end{equation}
\begin{align}
&a_1=1+\alpha \epsilon^2 (k^2+k'^2) \frac{1+\sin^2 \theta}{\cos^2\theta}+\frac{\alpha^2 \epsilon^4}{\cos^4\theta}
(k^2+ k'^2  \sin^2\theta) (k'^2+ k^2  \sin^2\theta)\\
&a_2=\frac{4\alpha \epsilon^2 \sin \theta}{\cos^2 \theta} \left(1+\alpha \epsilon^2 (k^2+k'^2) (\sec^2\theta-\frac{1}{2}) \right) k k'\\
&a_3=\frac{4\alpha^2 \sin^2\theta}{\cos^4\theta} k^2k'^2 \epsilon^4 \\ 
&a_4=\frac{2kk' \sin \theta \epsilon^2 }{\cos^2 \theta}\\
&a_5=\frac{k'^2+k^2+q^2\cos^2 \theta}{2kk'\sin \theta}
\end{align}
}

\section{Universal functions at unitarity}
\label{app:phi_unitary}

The recurrence relation satisfied by the universal functions ${\langle z | \phi_{\ell,\expos,\tau}\rangle}$ at unitarity i.e. when $\tau=0$ is detailed in the following lines.
%Using the Fourier transform, one obtains the function ${\langle k | \mathcal A_\ell \rangle}$ from ${\langle \rho | \mathcal A_\ell \rangle}$ in terms of the Hankel transform:
%\begin{equation}
%\langle k | \mathcal A_\ell \rangle = \int_0^\infty \rho^2 j_\ell(k\rho) \langle \rho |A_\ell \rangle d\rho .  
%\label{eq:Hankel_transform} 
%\end{equation}
%Universal functions at unitarity with the normalization ${\langle z | \phi_{l,\expos,\tau}\rangle \sim z^{\expos-2}}$ when ${z\to \infty}$, 
%and for higher values of ${\ell}$, one has the recurrence
%\begin{equation}
%\langle z| \phi_{\ell+1,\expos,0}\rangle =\frac{(\expos+\ell) \langle z| \phi_{\ell-1,\expos,0}\rangle }{\expos-\ell-1}+
%\frac{(2\ell+1) \langle z| \phi_{\ell,\expos+1,0}\rangle }{z(\ell+1-\expos)} \\
%+\frac{(2\ell+1)(\expos+\ell) \langle z| \phi_{\ell,\expos-1,0}\rangle}
%{4z\expos (1-\expos)} .
%\label{eq:recurrence_functions}
%\end{equation}
%\section{Recurrence relation of the universal functions at unitarity}
At unitarity, the wavefunction of total angular momentum ${\ell\hbar}$ is separable  in the configuration space as
\begin{equation}
\langle \boldsymbol \rho | \Psi \rangle =\frac{\langle \rho | \mathcal A_\ell\rangle}{\rho} \Phi_\ell(\boldsymbol \rho/\rho) .
\label{eq:separability}
\end{equation}
The function ${\Phi_\ell(\boldsymbol \rho/\rho)}$ in Eq.~\eqref{eq:separability} is an eigenstate of the Laplacian on the unit sphere of dimension 5 with the eigenvalue ${4-\expos^2}$. It satisfies the two-body Bethe-Peierls condition at the contact of each interacting pair $(ij)$ and incorporates the exchange symmetries of the configuration studied. The hyperradial function satisfies a 2D stationary Schr\"{o}dinger equation with a generalized centrifugal barrier which can be viewed as a pure inverse square potential:
\begin{equation}
\left(-\partial_\rho^2 -\frac{1}{\rho} \partial_\rho +\frac{\expos^2}{\rho^2} +q^2 \right) \left(\rho \langle \rho | \mathcal A_\ell\rangle\right)=0 .
\end{equation}
The solution of this last equation which is bounded for ${\rho \to \infty}$ is the Macdonald function up to a normalization constant:  
\begin{equation}
\langle \rho | \mathcal A_\ell\rangle = \mathcal N K_\expos(q\rho)/\rho .
\end{equation}
At unitarity, the function ${\langle z| \phi_{\ell,\expos,0}\rangle}$  is thus given by the Hankel transform
\begin{equation}
\langle z| \phi_{\ell,\expos,0}\rangle=  
\frac{1}{\mathcal N_{\ell,\expos}} \int_0^\infty  K_\expos(u)  j_\ell(zu) u du,
\label{eq:Hankel_phi_u}
\end{equation}
where the normalization constant ${\mathcal N_{\ell,\expos}}$ is such that ${\langle z| \phi_{\ell,\expos,0}\rangle \simeq z^{-2+\expos}}$ when ${z\to \infty}$. 
The universal functions for $\ell=1$ and $\ell=2$ the universal functions are then obtained directly from Eq.~(\ref{eq:Hankel_transform}) and the known expression in the configuration space in Eq.~(\ref{eq:Fradial}):
\begin{equation}
\langle z | \phi_{0,\expos,0}\rangle = 2^{1-\expos} \frac{\sinh\left[\expos\arcsinh (z)\right]}{z\sqrt{1+z^2}}  
\quad ; \quad
\langle z |\phi_{1,\expos,0}\rangle = \frac{2^{1-\expos}}{(\expos-1)} \frac{d}{dz} \left(\frac{\sinh\left[\expos\arcsinh (z)\right]}{z} \right) .
\label{eq:phi_1}
\end{equation}
%\begin{align}
%&\langle z | \phi_{0,\expos,0}\rangle = 2^{1-\expos} \frac{\sinh\left[\expos\arcsinh (z)\right]}{z\sqrt{1+z^2}} 
%\label{eq:phi_0}\\
%&\langle z |\phi_{1,\expos,0}\rangle = \frac{2^{1-\expos}}{(\expos-1)} \frac{d}{dz} \left(\frac{\sinh\left[\expos\arcsinh (z)\right]}{z} \right) .
%\label{eq:phi_1}
%\end{align}
For higher values of ${\ell}$, one uses the recurrence relations satisfied by the Bessel functions
\begin{equation}
j_{\ell+1}(zu)= (2\ell+1) \frac{j_\ell(zu)}{zu} -j_{\ell-1}(zu)\quad ; \quad  K_\expos(u) = \frac{u}{2\expos} \left( K_{\expos+1}(u) -  K_{\expos-1}(u) \right) .
\label{eq:recurrence_Bessel}
\end{equation}
This gives the recurrence relation for the normalization constant
\begin{equation}
\mathcal N_{\ell+1,\expos}=  \frac{(2\ell+1)}{2\expos} \mathcal N_{\ell,\expos+1}- {\mathcal N_{\ell-1,\expos}} .
\label{eq:recurrence_Norm}
\end{equation}
Using the expressions of ${\mathcal N_{\ell,\expos}}$ at ${\ell=0,1}$
\begin{equation}
\mathcal N_{0,\expos} = \frac{2^{\expos-2} \pi}{\sin \left( \frac{\pi \expos}{2} \right)} 
\quad \text{and} \quad
\mathcal N_{1,\expos} =  \frac{2^{\expos-2} \pi  (1-\expos)}{\expos \cos\left( \frac{\pi \expos}{2} \right)}, 
\label{eq:N_0_N_1}
\end{equation}
one finds from Eq.~\eqref{eq:recurrence_Norm}:
\begin{equation}
\mathcal N_{\ell,\expos}= \frac{2^{\expos-2}\pi}{\sin \left( \frac{\pi}{2} (\expos-\ell)\right)}
\times
\left\{ \begin{array}{lc}
\displaystyle \frac{(\expos-\ell)(\expos-\ell+2) \dots (\expos-2)} {(\expos+\ell-1)(\expos+\ell-3)\dots(\expos+1)} &\text{for} \  \ell \ \text{even}\\
\displaystyle \frac{(\expos-\ell)(\expos-\ell+2) \dots (\expos-1)}{(\expos+\ell-1)(\expos+\ell-3) \dots \expos} &\text{for} \  \ell \ \text{odd}.
 \end{array}
\right. 
\label{eq:Nell}
\end{equation}
Using again the relations in Eq.~\eqref{eq:recurrence_Bessel}, one obtains the recurrence for the universal functions at unitarity
\begin{equation}
\langle z| \phi_{\ell+1,\expos,0}\rangle =-\frac{\mathcal N_{\ell-1,\expos}}{\mathcal N_{\ell+1,\expos}} \langle z| \phi_{\ell-1,\expos,0}\rangle +
\frac{2\ell+1}{2\expos z \mathcal N_{\ell+1,\expos} } 
 \left( \mathcal N_{\ell,\expos+1}
\langle z| \phi_{\ell,\expos+1,0}\rangle 
-\mathcal N_{\ell,\expos-1} \langle z| \phi_{\ell,\expos-1,0}\rangle \right) \ \text{for} \ \ell\ge1 .
\label{eq:recurrence_functions}
\end{equation}
Finally, one can derive the expression of the function ${\langle z |\phi_{\ell,\expos,0}\rangle}$ for any desired values of ${\ell}$ from the expressions of ${\langle  z| \phi_{\ell,\expos,0}\rangle}$  in Eq.~\eqref{eq:phi_1} and the recurrence relation obtained from  Eq.~\eqref{eq:Nell} and Eq.~\eqref{eq:recurrence_functions}:
\begin{equation}
 \langle z| \phi_{\ell+1,\expos,0}\rangle =\frac{(\expos+\ell) \langle z| \phi_{\ell-1,\expos,0}\rangle }{{\expos-\ell-1}}+
\frac{(2\ell+1) \langle z| \phi_{\ell,\expos+1,0}\rangle }{(\ell+1-\expos)z} 
+\frac{(2\ell+1)(\expos+\ell) \langle z| \phi_{\ell,\expos-1,0}\rangle}
{4\expos (1-\expos)z} \ \text{for} \ \ell\ge1
  ,
\label{eq:recurrence_functions_bis}
\end{equation}

\section{Coefficients of the expansion of the universal functions up to the second order}
\label{app:coef_expansion}

Applying the Hankel transform in Eq.~\eqref{eq:Hankel_transform} on each term of Eq.~\eqref{eq:expansion_A}, one finds the following series representation of the universal functions:
\begin{equation}
\langle z|\phi_{\ell,\expos,\tau}\rangle = \sum_{n=0}^\infty b_{\ell,-\expos,\tau}^{(n)} z^{\expos-2-n}
 -   \frac{\Gamma(1+\ell+\expos)r_{\ell,\expos,\tau}}{\Gamma(1+\ell-\expos)}  \sum_{n=0}^\infty b_{\ell,\expos,\tau}^{(n)} z^{-\expos-2-n} .
\label{eq:series_phi}
\end{equation}
with the choice of normalization ${b_{\ell,\gamma,\tau}^{(0)}=1}$ and where the coefficient ${b_{\ell,\gamma,\tau}^{(n)}}$ is directly obtained from the coefficient ${c_{\ell,\gamma,\tau}^{(n)}}$ by using the known integral 
\begin{equation}
\int_0^\infty \rho^\beta j_\ell(k \rho) d\rho=\sqrt{\pi} \frac{2^{\beta-1}}{k^{\beta+1}} \frac{\Gamma\left(\frac{1+\ell+\beta}{2}\right)}{\Gamma\left(1+\frac{\ell-\beta}{2}\right)} .
\label{eq:Hankel_power_law_appendix}
\end{equation}
For instance when ${n=1,2}$ 
\begin{equation}
c_{\ell,\gamma,\tau}^{(1)}=
\frac{\Gamma\left(1+\frac{\ell+\gamma}{2}\right) \Gamma\left(\frac{\ell-\gamma}{2}\right) b_{\ell,\gamma,\tau}^{(1)}}
{2\Gamma\left(\frac{1}{2} +\frac{\ell-\gamma}{2}\right) \Gamma\left(\frac{3}{2}+\frac{\ell+\gamma}{2}\right)} 
\quad ; \quad
c_{\ell,\gamma,\tau}^{(2)}= \frac{b_{\ell,\gamma,\tau}^{(2)}}{(2+\ell+\gamma)(\ell-\gamma-1)}  {.}
\label{eq:b2c}
\end{equation}
In Eq.~\eqref{eq:b2c} and in all subsequent equations ${\gamma=\pm \expos}$.

Each sum in the right hand side of Eq.~\eqref{eq:series_phi} is a solution of the STM equation but which is unbounded for $z\to0$. In the same spirit as the Frobenius method, one injects one of the sum in Eq.~\eqref{eq:STM_vp} divided by $z$. Then identification of each term order by order in powers of ${1/z}$ is a way to obtain all the coefficients of the series: 
\begin{multline}
 \left(\frac{\tau}{z}-1-\frac{1}{2z^2}+\frac{1}{8z^4}\dots \right) \left( 1+ \frac{b_{\ell,\gamma,\tau}^{(1)}}{z}+ \frac{b_{\ell,\gamma,\tau}^{(2)}}{z^2}+\dots  
\right)+  \frac{2\stat(-1)^\ell}{\sin (2\theta)}
\Biggl[ \int_0^\infty \frac{dv}{\pi} Q_\ell\left(\frac{1+v^2}{2v\sin \theta}\right) v^{-\gamma-1}
\\ \times \left(1+\frac{b_{\ell,\gamma,\tau}^{(1)}}{z v }+ \dots\right) +\frac{\cos^2 \theta}{2\sin \theta}
\int_0^\infty \frac{dv}{\pi} Q_\ell^{(1)}\left(\frac{1+v^2}{2v\sin \theta}\right) \frac{v^{-\gamma-2}}{z^2} 
 \left(1+\frac{b_{\ell,\gamma,\tau}^{(1)}}{z v }
+ \frac{b_{\ell,\gamma,\tau}^{(2)}}{z^2v^2}+
\dots \right)  +\dots \Biggr]=0 
 \label{eq:STM_series}
\end{multline}
where ${Q_\ell^{(n)}(t)=\frac{d^n}{dt^n}Q_\ell(t)}$. One defines the function
\begin{equation} 
I^{(n)}_{\ell,\gamma}=(-1)^\ell \int_0^\infty \frac{dv}{\pi}  Q_\ell^{(n)}\left(\frac{v^2 + 1}{2 v \sin \theta}\right)  v^{\gamma-1}
\label{eq:int_gamma}
\end{equation}
which satisfies ${I^{(n)}_{\ell,\gamma}=I^{(n)}_{\ell,-\gamma}}$. It can be calculated explicitly for ${\ell=0,1}$ and for larger values of ${\ell}$, one uses
\begin{align}
&(2\ell+1) t Q_\ell(t)=(\ell+1) Q_{\ell+1}(t)+\ell Q_{\ell-1}(t) \\
&Q_{\ell+1}^{(1)}(t) =\frac{2\ell+1}{\ell+1} \left[ Q_\ell(t) + t Q_\ell^{(1)}(t) \right] - \frac{\ell Q_{\ell-1}^{(1)}(t)}{\ell+1} .
\end{align}
For instance the recurrence of ${I^{(0)}_{\ell,\gamma}}$ and ${I^{(1)}_{\ell,\gamma}}$ are:
\begin{align}
&I_{\ell+1,\gamma}^{(0)} =- \frac{(2\ell +1)  \left( I_{\ell,\gamma-1}^{(0)} +I_{\ell,\gamma+1}^{(0)} \right)}{2 (\ell+1) \sin \theta} 
-\frac{\ell I_{\ell-1,\gamma}^{(0)}}{\ell+1}\label{eq:I0_recurrence}\\
&I^{(1)}_{\ell+1,\gamma}=-\frac{2\ell+1}{\ell+1}
\left(I_{\ell,\gamma}^{(0)}+\frac{I^{(1)}_{\ell,\gamma+1}+I^{(1)}_{\ell,\gamma-1}}{2 \sin \theta}\right)
-\frac{\ell}{\ell+1} I^{(1)}_{\ell-1,\gamma} 
\end{align} 
with the initial conditions ${I^{(0)}_{-1,\gamma} =I^{(1)}_{-1,\gamma} =0}$ and
\begin{equation}
I^{(0)}_{0,\gamma}= \frac{ \sin ( \gamma \theta) }{\gamma \cos ( \gamma \pi/2)}  \ ; \ 
I^{(1)}_{0,\gamma}= -\frac{\sin(\gamma \theta)\tan \theta}{\sin (\pi \gamma/2)} .
\end{equation}
At zero order, Eq.~\eqref{eq:STM_series} gives the equation satisfied by ${\gamma}$:
\begin{equation}  
{ \sin 2 \theta}  = {2 \stat}  I^{(0)}_{\ell,\gamma} 
\label{eq:gamma}
\end{equation}
and at first up to third order in ${(1/z)}$, one obtains
\begin{align}
&b_{\ell,\gamma,\tau}^{(1)}=\frac{\tau I^{(0)}_{\ell,\gamma}}{I^{(0)}_{\ell,\gamma}-I^{(0)}_{\ell,1+\gamma}} ,
\label{eq:b1}\\ 
&b_{\ell,\gamma,\tau}^{(2)}=\frac{2 \tau I^{(0)}_{\ell,\gamma}b_{\ell,\gamma,\tau}^{(1)}
+\frac{\cos^2\theta}{\sin \theta} I^{(1)}_{\ell,1+\gamma}-I^{(0)}_{\ell,\gamma}}{2\left(I^{(0)}_{\ell,\gamma}-I^{(0)}_{\ell,2+\gamma}\right)} ,
\label{eq:b2}\\
&b_{\ell,\gamma,\tau}^{(3)}
=\frac{2\tau I^{(0)}_{\ell,\gamma} b_{\ell,\gamma,\tau}^{(2)}
+\left(
\frac{  \cos^2\theta}{\sin \theta} I^{(1)}_{\ell,2+\gamma}-I^{(0)}_{\ell,\gamma} \right) b_{\ell,\gamma,\tau}^{(1)}}
{ 2\left(I^{(0)}_{\ell,\gamma} - I^{(0)}_{\ell,3+\gamma}\right)} .
\label{eq:b3}
\end{align}
At higher orders, for even (odd) values of ${n=2p}$ (${n=2p+1}$) , ${b_{\ell,\gamma,\tau}^{(2p)}}$ and ${\left( b_{\ell,\gamma,\tau}^{(2p+1)}/\tau\right)}$ are polynoms of ${\tau^2}$ of order $p$. One can verify that the expression of the coefficient ${b_{\ell,\gamma,0}^{(2)}}$ in Eq.~\eqref{eq:b2} coincides with the result obtained from the expansion of ${K_\expos(q\rho)}$ and Eq.~\eqref{eq:b2c}. For instance one has for ${\ell=0,1}$:
\begin{equation}
b_{0,\gamma,0}^{(2)} = \frac{-\gamma-2}{4} \quad  ; \quad b_{1,\gamma,0}^{(2)} = \frac{-\gamma(3+\gamma)}{4 (1+\gamma)}  .
\label{eq:b01_unitary}
\end{equation}
Finally one obtains 
\begin{align}
&c_{\ell,\gamma,\tau}^{(1)} =
\frac{\tau I^{(0)}_{\ell,\gamma} \Gamma\left(1+\frac{\ell+\gamma}{2}\right) \Gamma\left(\frac{\ell -\gamma}{2}\right)}{2 \left(I^{(0)}_{\ell,\gamma}-I^{(0)}_{\ell,1+\gamma}\right) \Gamma\left(\frac{1+\ell -\gamma}{2}\right) \Gamma\left(\frac{3+\ell+\gamma}{2}\right)}
\label{eq:c1} \\
&c_{\ell,\gamma,\tau}^{(2)} =\frac{1}{4+4\gamma} + \frac{\left( \tau I^{(0)}_{\ell,\gamma}\right)^2}{(\ell+2+\gamma)(\ell-1-\gamma)(I^{(0)}_{\ell,\gamma}-I^{(0)}_{\ell,1+\gamma})(I^{(0)}_{\ell,\gamma}-I^{(0)}_{\ell,2+\gamma})}
\label{eq:c2} .
\end{align}
One has also from the series representation of ${K_\expos(q\rho)}$:
\begin{equation}
r_{\ell,\expos,0}=\frac{\pi 4^{-\expos}}{\expos \Gamma(\expos)^2 \sin(\pi \expos)} .
\label{eq:r_ls0}
\end{equation}
Finding {an} analytical expression for the coefficient ${r_{\ell,\expos,\tau}}$ away from unitarity, i.e. when ${\tau}$ is not zero, remains an unsolved issue. Nevertheless, in the limit of a large and negative value of ${\tau}$, one has ${r_{\ell,\expos,\tau}=O(|\tau|^{2\expos})}$. This property plays an important role for the endpoint of the spectrum at negative scattering length. A proof can be obtained by considering the STM equation at zero energy and for a negative scattering length with the change of variable ${v=k|\asc|}$:
\begin{equation}
\frac{2\stat (-1)^\ell }{\pi \sin (2 \theta)} \int_0^\infty \frac{v'dv'}{v}  
 Q_\ell\left(\frac{v^2+v'^2}{2 v v' \sin \theta } \right) \langle v' |\mathcal A_\ell \rangle 
= \langle v |\mathcal A_\ell \rangle (v+1) .
\label{eq:STM_tau_infty}
\end{equation}
In the limit of large values of $v$, one recovers the two independent solutions ${v^{-2\pm \expos}}$ and the general solution is of the form
\begin{equation}
\langle v | \mathcal A_\ell \rangle = v^{\expos-2} \left(1+ \sum_{n=1}^\infty a_{\ell,-\expos}^{(n)} v^{-n} \right)
-\frac{\Gamma(1+\ell+\expos)B}{\Gamma(1+\ell-\expos)} v^{-\expos-2} \left(1+ \sum_{n=1}^\infty a_{\ell,\expos}^{(n)} v^{-n} \right)  .
\label{eq:series_tau_infty}
\end{equation}
The expansion of Eq.~\eqref{eq:series_tau_infty} is injected in Eq.~\eqref{eq:STM_tau_infty} successively for ${B=0}$ and ${B=\infty}$ to obtain the relations satisfied by the coefficient ${a_{\ell,- \expos}^{(n)}}$ and ${a_{\ell,\expos}^{(n)}}$. Comparison with Eq.~\eqref{eq:STM_series}  in the limit ${\tau\to-\infty}$ and ${z/|\tau|\ll1}$ permits one to show that ${a_{\ell,\pm \expos}^{(n)} \sim b_{\ell,\pm \expos,\tau}^{(n)}/|\tau|^n}$. Using the identity ${v=z/|\tau|}$ in Eq.~\eqref{eq:series_tau_infty}, and from the comparison with Eq.~\eqref{eq:series_phi}, one obtains ${r_{\ell,\expos,\tau} \sim B |\tau|^{2\expos}}$ where $B$ is a function of ${\stat}$, ${\theta}$ and ${\ell}$.

\section{Computation of the universal functions}
\label{app:computation}
 {The method used for the numerical evaluation of the universal functions and of the coefficient ${r_{\ell,\expos,\tau}}$ when ${\tau\ne0}$ with the assumption that ${\expos<2}$ is detailed in this Appendix.} This method can be thus applied for all the ITBR of the 2FI systems when ${\ell=1}$. In the numerical analysis, the grid for the variable ${z\in \{z_1\dots z_N\}}$ is defined by an exponential change of variable ${z_n=z_0 e^{(n-1)h}}$ where ${h\ll 1}$ which permits one to sample the function on a rather large interval. In the calculations, typically from 1000 to 2000 points have been used with ${z_0=10^{-3}}$ {and ${z_N}$ from ${10^9}$ to ${10^{13}}$}. The universal functions are discretized with
\begin{equation}
S_n(\tau)=z_n^{3/2} \langle z_n| \phi_{\ell,\expos,\tau}\rangle  .
\end{equation}
where the prefactor makes the linear system symmetrical. Using the Lippmann Schwinger-like equation, the vector ${S_n(\tau)}$ is obtained from ${S_n(0)}$ by means of a matrix inversion with
\begin{equation}
S_n(\tau)=S_n(0)+\delta S_n(\tau) \quad \text{and} \quad \delta S_n(\tau) = \tau \sum_{n'} M_{n,n'}^{-1} S_{n'}(0) 
\end{equation}
where the matrix ${M_{n,n'}}$ is given by:
\begin{equation}
M_{n,n'} =(\sqrt{z_{n}^2 +1} -\tau) \delta_{n,n'} -  \frac{2 h \stat }{\pi \sin 2 \theta} (-1)^\ell
\sqrt{z_{n} z_{n'}} Q_{\ell} \left(\frac{z_{n}^2 +z_{n'}^2+\cos^2\theta}{2  z_{n} z_{n'}\sin \theta} \right) .
\end{equation}
The universal functions are then obtained numerically by
\begin{equation}
\langle z_n | \phi_{\ell,\expos,\tau}\rangle =  \langle z_n | \phi_{\ell,\expos,0}\rangle + \langle z_n |\delta \phi_{\ell,\expos,\tau}\rangle \ \text{with} \   \langle z_n |\delta \phi_{\ell,\expos,\tau}\rangle =\delta S_n(\tau) z_n^{-3/2} .
\end{equation}
Depending on the value of $\expos$, the dominant terms in the expansion of ${ \langle z_n |\delta \phi_{\ell,\expos,\tau}\rangle }$ for large ${z_n}$ are not the same. One then uses the following ansatz for ${\langle z_n |\delta \phi_{\ell,\expos,\tau}\rangle}$:
\begin{equation}
 \langle z_n |\delta \phi_{\ell,\expos,\tau}\rangle=
\left\{
\begin{array}{ll}
 z_n^{\expos-2} \left(\sum_{i=1}^2 \frac{ \delta b^{(i)}_{\ell,-\expos,\tau}(n)}{(z_n)^i} \right)
-  z_n^{-\expos-2}  \frac{\Gamma(1+\ell-\expos)}{\Gamma(1+\ell+\expos) } \delta r_{\ell,\expos,\tau}(n) 
\left(1 + \sum_{i=1}^2 \frac{ \delta b^{(i)}_{\ell,\expos,\tau}(n)}{(z_n)^i} \right)   
& \text{for} \ \expos\le \frac{1}{2}\\
 z_n^{\expos-2} \left( \sum_{i=1}^3 \frac{ \delta b^{(i)}_{\ell,-\expos,\tau}(n)}{(z_n)^i} \right)
-  z_n^{-\expos-2} \frac{\Gamma(1+\ell-\expos)}{\Gamma(1+\ell+\expos) } \delta r_{\ell,\expos,\tau}(n) 
\left(1 + \frac{ \delta b^{(1)}_{\ell,\expos,\tau}(n)}{z_n}
\right)  & \text{for} \ \frac{3}{2}\ge\expos>\frac{1}{2}\\
 z_n^{\expos-2} \left(\sum_{i=1}^4 \frac{ \delta b^{(i)}_{\ell,-\expos,\tau}(n)}{(z_n)^i} \right)
-  z_n^{-\expos-2}  \frac{\Gamma(1+\ell-\expos)}{\Gamma(1+\ell+\expos) } \delta r_{\ell,\expos,\tau}(n) & \text{for} \ 2>\expos>\frac{3}{2}.
\end{array}
\right.
\label{eq:discretized_Sn}
\end{equation}
When for large values of ${n}$, the functions ${[\delta r_{\ell,\expos,\tau}(n),\delta b_{\ell,\gamma,\tau}^{(i)}(n)]}$ are constant, Eq.~\eqref{eq:discretized_Sn} coincides with the first five dominant terms of the expansion of the function ${\langle z |\delta  \phi_{\ell,\expos,\tau}\rangle}$ in the limit ${z\gg1}$. For higher values of the scaling exponent, more terms in the ansatz have to be added, however the unavoidable numerical noise makes this numerical method ineffective. The functions ${[\delta r_{\ell,\expos,\tau}(n),\delta b_{\gamma,\tau}^{(i)}(n)]}$ are obtained by solving a ${5 \times 5}$ linear system deduced from the expression at a given $n$ of the first four discretized log-derivatives in ${\left(z\frac{d}{dz}\right)}$  of the two sides of Eq.~\eqref{eq:discretized_Sn} where ${[\delta r_{\ell,\expos,\tau}(n),\delta b_{\ell,\gamma,\tau}^{(i)}(n)]}$ are considered as constant. Precise results are found by using a discretization of the derivatives with nine points. One searches for plateaus in the functions ${[\delta r_{\ell,\expos,\tau}(n),\delta b_{\ell,\gamma,\tau}^{(i)}(n)]}$  for a sufficiently large value of ${n}$ such that ${z_n\gg1}$, up to a value of ${n}$ where the numerical signal is too noisy, i.e. in the far tail of the function ${ \langle z_n |\delta \phi_{\ell,\expos,\tau}\rangle}$. The value ${\delta b_{\ell,\gamma,\tau}^{(i)}(n)}$ (or ${\delta r_{\ell,\expos,\tau}(n)}$) corresponding to a minimum of slope on the plateau (whenever it exists) is denoted by $\delta b_{\ell,\gamma,\tau}^{(i),{\rm as}}$ (or ${\delta r_{\ell,\expos,\tau}^{\rm as}}$). This numerical procedure permits a numerical evaluation of the corresponding coefficients in the expansion of ${\langle z|\phi_{\ell,\expos,\tau}\rangle}$ at large ${z}$:
\begin{equation}
b_{\ell,\gamma,\tau}^{(i)} = b_{\ell,\gamma,0}^{(i)} + \delta b_{\ell,\gamma,\tau}^{(i),{\rm as}}\  \text{and}
\quad r_{\ell,\expos,\tau}=r_{\ell,\expos,0} + \delta r_{\ell,\expos,\tau}^{\rm as} .
\end{equation}
The parameter ${r_{\ell,\expos,\tau}}$ is the only universal quantity which is obtained from a numerical study of the dimensionless STM equation. The precision of this method can be tested by considering the term ${b_{\ell,-\expos,\tau}^{(i)}(n) z^{\expos-2-n}}$ corresponding to the subdominant contribution with respect to the term in ${r_{\ell,\expos,\tau} z^{-\expos-2}}$: ${\delta b_{\ell,-\expos,\tau}^{(1)}(n)}$ for ${\expos<1/2}$, ${\delta b_{\ell,-\expos,\tau}^{(2)}(n)}$ for ${1/2<\expos<1}$, ${\delta b_{\ell,-\expos,\tau}^{(3)}(n)}$ for ${1<\expos<3/2}$ \dots Comparing the value at the plateau with respect to the exact asymptotic value ${b_{\ell,-\expos,\tau}^{(n)}}$ given in Appendix \ref{app:coef_expansion} permits one to evaluate the precision made on the value of ${r_{\ell,\expos,\tau}}$. Examples of computations are given in Figs.~\ref{fig:deltab_1}-\ref{fig:deltab_4}. From this basis, the typical relative precision for ${r_{\ell,\expos,\tau}}$ is of the order of  ${10^{-3}}$ or much smaller in the interval ${0<\expos<3/2}$. For larger values of ${\expos}$, the plateau of ${\delta r_{\ell,\expos,\tau}(n)}$ obtained numerically is less pronounced and thus the evaluation of ${r_{\ell,\expos,\tau}}$ is not very precise. This is not a problem for the determination of the binding energy from Eq.~\eqref{eq:energy-condition} where ${r_{\ell,\expos,\tau}}$ is needed only for ${\expos<1.5}$.
\begin{figure}
\centering{\includegraphics[width=10cm]{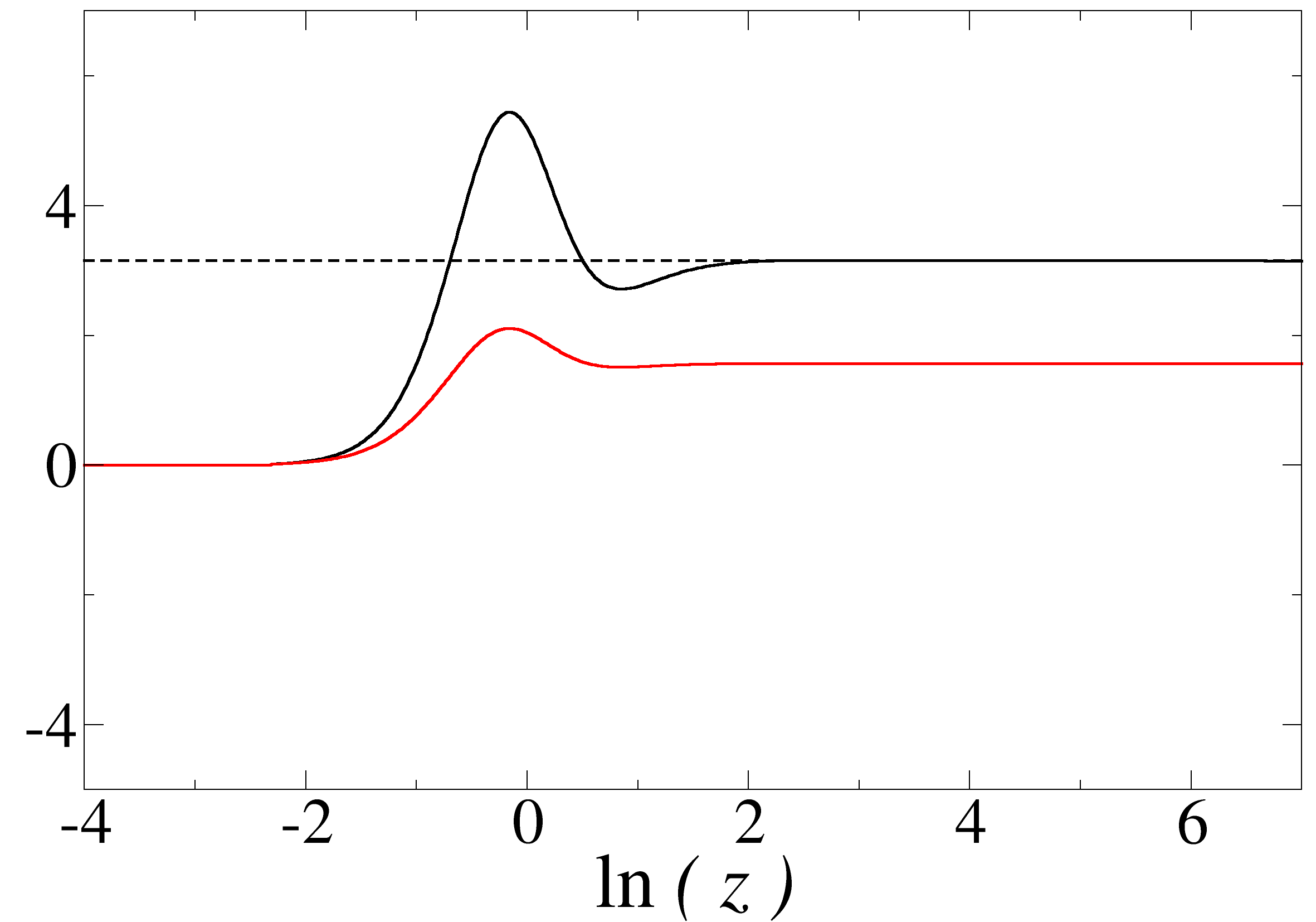}}
\caption{Black solid line: plot of the function ${\delta b^{(1)}_{\ell,-\expos,\tau}(n)}$ for ${\ell=1,\expos=1/4,\tau=-1/2}$; dotted line: exact value when $n\gg1$; red solid line: plot of ${\delta r_{\ell,\expos,\tau}(n)}$.}
\label{fig:deltab_1}
\end{figure}
\begin{figure}
\centering
\includegraphics[width=10cm]{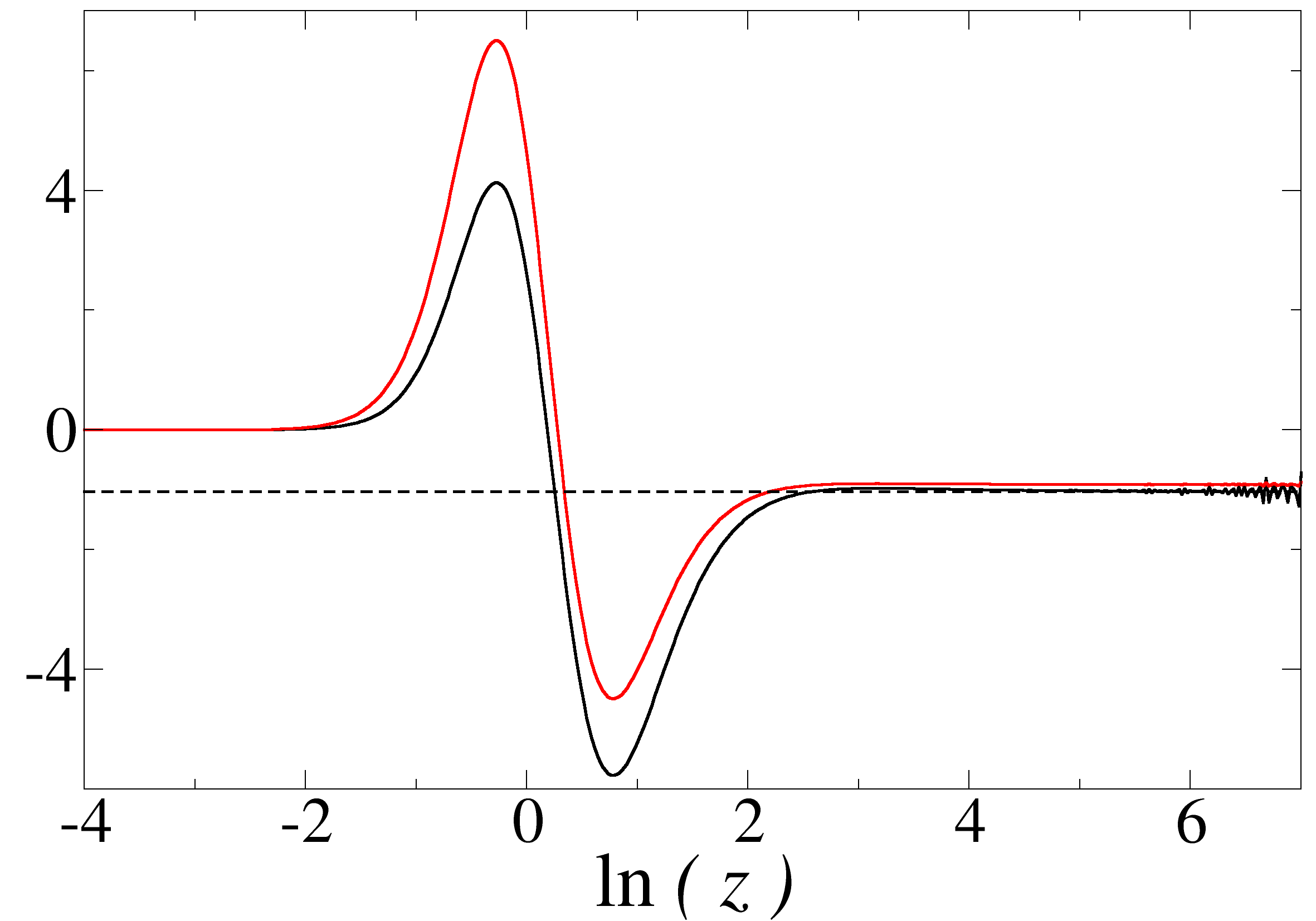}
\caption{Black solid line: plot of the function ${\delta b^{(3)}_{\ell,-\expos,\tau}(n)}$  for ${\ell=1,\expos=4/3,\tau=-1/2}$; dotted line: exact value when ${n\gg1}$; red solid line: plot of ${\delta r_{\ell,\expos,\tau}(n)}$.}
\label{fig:deltab_3}
\end{figure}
\begin{figure}
\centering
\includegraphics[width=10cm]{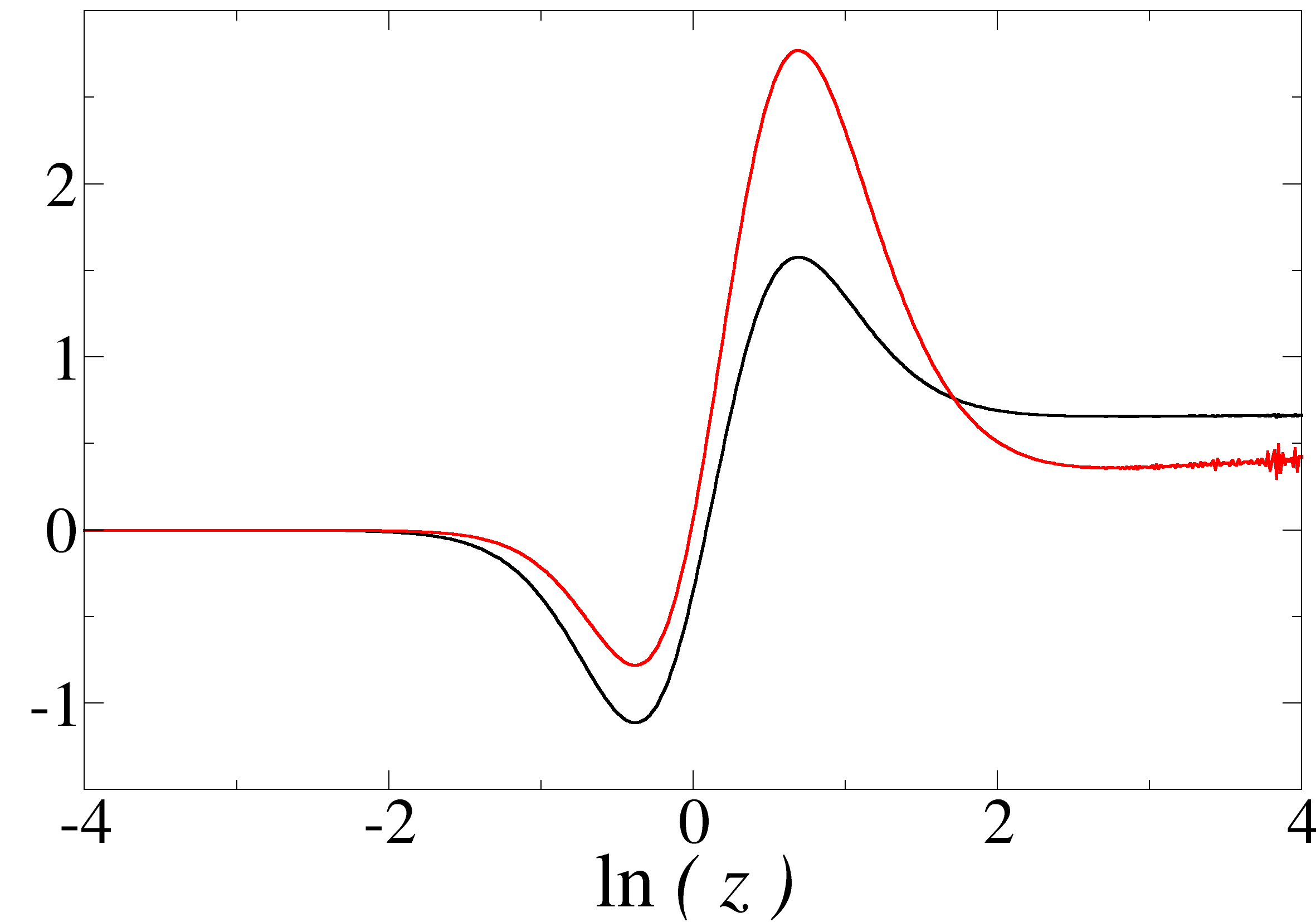}
\caption{Left figure: plot of the function ${\delta b^{(4)}_{\ell,-\expos,\tau}(n)}$  for ${\ell=1,\expos=1.8,\tau=-1/2}$;  red solid line: plot of ${\delta r_{\ell,\expos,\tau}(n)}$. The plateau of ${\delta r_{\ell,\expos,\tau}(n)}$ is less pronounced than the one obtained when ${\expos<3/2}$.}
\label{fig:deltab_4}
\end{figure}
One can notice that it is also possible to find an approximate numerical solution of the universal functions (nevertheless with a smaller accuracy than with the previous method) by imposing 'by hand' a remainder of the STM equation at a finite but large ${\Lambda}$ with the substitution ${\langle z |  \mathcal L^\infty - \mathcal L^\Lambda  |\phi_{\ell,\expos,\tau} \rangle\equiv 1}$ when ${z\simeq \Lambda}$ and zero otherwise. This term which plays the role of a source term mimics the effect of the actual short range interactions in a reference model where a resonant mode exists.
\end{widetext}

\begin{thebibliography}{99}
\bibitem{limit} {One has to distinguish the limit of ${\mathcal L^{\Lambda}}$ when ${\Lambda \to \infty}$ and ${\mathcal L^{\infty}}$ because ${\lim_{\Lambda \to \infty} \mathcal L^{\Lambda}  |\phi_{\ell,\expos,0} \rangle \ne \mathcal L^{\infty} |\phi_{\ell,\expos,0} \rangle=0}$.}
\end{thebibliography}
\end{document}